\documentclass[acmsmall,screen]{acmart}
\usepackage{threeparttable}
\usepackage{stfloats}
\usepackage{cases}
\usepackage{algorithm}
\usepackage{algorithmicx}
\usepackage{algpseudocode}
\usepackage{array}
\usepackage{url}
\usepackage{color}
\usepackage{subfigure}
\usepackage{verbatim}
\usepackage{multicol}
\usepackage{multirow}
\usepackage{booktabs}
\usepackage{makecell}
\usepackage{upgreek}

\AtBeginDocument{%
  }

\setcopyright{acmcopyright}
\copyrightyear{2025}
\acmYear{2025}
\acmDOI{XXXXXXX.XXXXXXX}

\acmJournal{TOSN}
\acmVolume{20}
\acmNumber{4}
\acmArticle{1}
\acmMonth{1}

\begin{document}

\title{Model-Driven Deep Neural Network for Enhancing Direction Finding with Commodity 5G gNodeB}

\author{Shengheng~Liu}
\authornote{Corresponding authors}
\orcid{0000-0001-6579-9798}
\affiliation{%
  \institution{National Mobile Communications Research Laboratory, Southeast University}
  \city{Nanjing}
  \country{China}
  \postcode{210096}}
\affiliation{%
  \institution{Purple Mountain Laboratories}
  \city{Nanjing}
  \country{China}
  \postcode{211111}}
\email{s.liu@seu.edu.cn}

\author{Zihuan~Mao}
\authornotemark[1]
\orcid{0000-0002-0661-2281}
\affiliation{%
  \institution{ZTE Corporation}
  \city{Nanjing}
  \country{China}
  \postcode{210012}}
\affiliation{%
  \institution{School of Information Science and Engineering, Southeast University}
  \city{Nanjing}
  \country{China}
  \postcode{210096}}
\email{mzh@seu.edu.cn}

\author{Xingkang~Li}
\orcid{0009-0002-1162-8349}
\affiliation{%
  \institution{School of Information Science and Engineering, Southeast University}
  \city{Nanjing}
  \country{China}
  \postcode{210096}}
\email{xingkangli@seu.edu.cn}

\author{Mengguan~Pan}
\orcid{0000-0001-5835-4539}
\affiliation{%
  \institution{School of Electronic and Information Engineering, Anhui University}
  \city{Hefei}
  \country{China}
  \postcode{230601}}
\affiliation{%
  \institution{Purple Mountain Laboratories}
  \city{Nanjing}
  \country{China}
  \postcode{211111}}
\email{panmengguan@ahu.edu.cn}

\author{Peng~Liu}
\orcid{0000-0001-8694-7091}
\affiliation{%
  \institution{Purple Mountain Laboratories}
  \city{Nanjing}
  \country{China}
  \postcode{211111}}
\email{liupeng@pmlabs.com.cn}

\author{Yongming~Huang}
\authornotemark[1]
\orcid{0000-0003-3616-4616}
\affiliation{%
  \institution{National Mobile Communications Research Laboratory, Southeast University}
\city{Nanjing}
\country{China}
\postcode{210096}}
\affiliation{%
  \institution{Purple Mountain Laboratories}
  \city{Nanjing}
  \country{China}
  \postcode{211111}}
\email{huangym@seu.edu.cn}

\author{Xiaohu~You}
\orcid{0000-0002-0809-8511}
\affiliation{%
  \institution{National Mobile Communications Research Laboratory, Southeast University}
\city{Nanjing}
\country{China}
\postcode{210096}}
\affiliation{%
  \institution{Purple Mountain Laboratories}
  \city{Nanjing}
  \country{China}
  \postcode{211111}}
\email{xhyu@seu.edu.cn}

\renewcommand{\shortauthors}{S. Liu et al.}

\begin{abstract}
Pervasive and high-accuracy positioning has become increasingly important as a fundamental enabler for intelligent connected devices in mobile networks. Nevertheless, current wireless networks heavily rely on pure model-driven techniques to achieve positioning functionality, often succumbing to performance deterioration due to hardware impairments in practical scenarios. Here we reformulate the direction finding or angle-of-arrival (AoA) estimation problem as an image recovery task of the spatial spectrum and propose a new model-driven deep neural network (MoD-DNN) framework. The proposed MoD-DNN scheme comprises three modules: a multi-task autoencoder-based beamformer, a coarray spectrum generation module, and a model-driven deep learning-based spatial spectrum reconstruction module. Our technique enables automatic calibration of angular-dependent phase error thereby enhancing the resilience of direction-finding precision against realistic system non-idealities. We validate the proposed scheme both using numerical simulations and field tests. The results show that the proposed MoD-DNN framework enables effective spectrum calibration and accurate AoA estimation. To the best of our knowledge, this study marks the first successful demonstration of hybrid data-and-model-driven direction finding utilizing readily available commodity 5G gNodeB.
\end{abstract}



\received{12 October 2023}

\maketitle

\section{Introduction}
\label{sec:intro}

High-accuracy positioning holds promise in supporting a diverse range of services within the forthcoming-generation of wireless communications \cite{YOU2022Toward}. The accurate position aids mobile devices in optimizing resource allocation and network configuration, thereby reducing communication latency and energy consumption \cite{wang2023eliminating}. Additionally, location awareness assumes a critical stance in bolstering security and privacy protection, facilitating identity authentication and safeguarding sensitive data in the sphere of cyber-physical systems. The advent of high-accuracy positioning services will transcend the conventional boundaries of mobile user lifestyles and usher in transformative prospects across diverse niche markets \cite{Dang2020What}. While a variety of techniques are available to accomplish this goal, including the global positioning system (GPS) \cite{Nguyen2020Multi}, ultra wide-band (UWB) \cite{pannuto2018harmonium}, and vision-inertial fusion \cite{teng2019cloudnavi}, each manifests distinct advantages and disadvantages. Nevertheless, none singularly achieves comprehensive coverage across both indoor and outdoor scenarios without the need for supplementary device installation. Recent developments have witnessed a notable surge in the deployment of fifth-generation (5G) base stations (a.k.a. gNodeBs or gNBs) in both indoor and outdoor environments to meet the demands for service quality and extensive coverage \cite{Li2022Mobility}, with their numbers projected to continue growing in the foreseeable future. This presents a viable opportunity to realize high-precision positioning by leveraging the 5G infrastructure itself, thereby obviating the necessity for additional hardware \cite{Pan2022Efficient}. Consequently, the integration of high data-rate communication and high accuracy positioning assumes an unprecedentedly heightened significance \cite{Liu2022Integrated}.

The positioning methods has been extensively explored and can be broadly categorized into two main groups: fingerprint-based methods and geometric-based methods. The fingerprint-based methods capitalizes on the uniqueness of the wireless channel that interconnects users and access point (AP), which is shaped by the intricate scattering environments \cite{Lin2021SateLoc}. However, the dynamic nature of the propagation medium introduces a discrepancy between the receive signal strength indicator (RSSI) and the outdated fingerprint database, thereby compromising the robustness of fingerprint-based methods \cite{Zhu2023BLS}. To overcome this issue, various strategies have been proposed that involve fusing derivative fingerprints to extract the resilient component of RSSI and enhance the accuracy of positioning \cite{Guo2020Robust}. However, the computational complexity of these methods grows exponentially with the expansion of the database,  thus posing a formidable challenge in terms of the requisite computation resources \cite{Huang2020Machine}. In contrast to the fingerprint-based methods, geometric-based methods impose minimal requirements concerning environmental dynamics \cite{Chen2022Carrier}. The 3rd Generation Partnership Project (3GPP) Release 16 \cite{LCS3GPP2021}, which reached its finalization in 2021, introduced several geometric-based schemes tailored for 5G. These encompass downlink-time-difference-of-arrival (DL-TDoA), uplink-time-difference-of-arrival (UL-TDoA), downlink-angle-of-departure (DL-AoD), uplink-angle-of-arrival (UL-AoA), and multi-round trip time (Multi-RTT) positioning techniques. Direction finding is a frequently visited problem in the realm of multichannel signal processing. When assuming an idealized mathematical model with perfect system response, maximum-likelihood (ML) \cite{Stoica1989ML} and subspace algorithms \cite{Schmidt1986Multiple} can be harnessed to estimate geometric parameters with super-resolution. However, the true system response invariably deviates from the highly idealized mathematical model. For instance, the accuracy of array response for angle-of-arrival (AoA) estimation is affected by mutual coupling effects, sensor location perturbations, gain-phase mismatches in multichannel receivers, and other unpredictable factors \cite{Liu2018Direction, Wang2018MAAD}. To tackle this challenge, the nonlinear fitting capabilities of artificial intelligence (AI) are envisioned to usher in a new paradigm for positioning \cite{Xiao2019Milli}.

The concept of AI-based direction finding was initially introduced by Rastogi et al. \cite{Rastogi1987Array}, where a Hopfield neural network was employed to establish a relationship between an energy function and signal parameters for direction estimation. More recent advancements have focused on deep neural network (DNN) approaches for AoA estimation in wireless systems \cite{Barthelme2021Machine}. One such approach involves a modified DNN with multiple hidden layers designed to extract sparse angular domain features, thereby enhancing the accuracy of estimation \cite{Huang2018Deep}. In parallel, feature-to-feature methodologies have been explored, offering incremental improvements in the physical interpretability of these models \cite{Xiang2020Improved}. By leveraging neural networks, specific input parameter features \cite{Naseri2022Machine} have been used for AoA estimation, which reinforces the linkage between input data and AoA estimation with clearer physical significance. Nevertheless, their accuracy relies heavily on the availability of extensive datasets, curtailing their ability to generalize effectively across diverse scenarios. A two-stage multi-layer perceptron was later proposed to improve AoA resolution and strengthen generalization capability \cite{ZhangTWO2024}. Still, these methods lack explicit designs to address hardware impairments, resulting in degraded estimation accuracy as such impairments become more influential.

To tackle this challenge, super-resolution AoA estimation was achieved by using time-domain signals as input to train the DNN \cite{ChenSDOA2024}.  Nonetheless, purely data-driven approaches like this focus on training a single parameter, often neglecting valuable information linked to other critical feature parameters \cite{Jiang2019Joint}. In contrast to neural networks that directly output AoA values, some methods solve the problem indirectly by reconstructing the spatial spectrum \cite{Wu2019Deep, Ren2022GoPose}. Adaptations of the multiple signal classification (MUSIC) algorithm through deep learning have demonstrated superior spectrum recovery, particularly under low signal-to-noise ratio (SNR) conditions and limited snapshots \cite{Elbir2020DeepMUSIC, Merkofer2022Deep}. MUSIC, a well-established model-driven method, benefits from integrating physical constraints within data-driven pipelines, enhancing interpretability and generalizability while reducing training costs. Additionally, the inherent model mismatch error, a typical limitation of model-driven approaches, can be mitigated with such hybrid frameworks \cite{kadambi2023incorporating}. A similar model-driven deep learning approach was recently introduced to reconstruct an ideal covariance matrix for root-MUSIC algorithm \cite{DuyUNet2024}. However, extracting features from complex manifold data remains a challenge, and the root-MUSIC algorithm still requires eigendecomposition with high complexity.

Motivated by the aforementioned facts, in this paper, we present a novel framework termed the model-driven deep neural network (MoD-DNN) framework for UL-AoA estimation employing a commodity 5G gNodeB while accounting for the widespread existence of hardware impairments, which bridges a critical gap in the field. While there have been a few pioneering efforts on 5G positioning \cite{yan2022privacy, Chen2022Carrier}, they are still at a very early stage of concept validation, which relies on software-defined radio (SDR) platforms for transmitting and receiving simulated 5G signals. Specifically, we first design a frequency diverse multi-task autoencoder for spatial beamforming. Through encoding and multi-task decoding, the input signal is filtered into different adjacent subregions, thereby enhancing the consistency of hardware impairments within each subregion. In the coarray spectrum generation module, we transform the covariance matrix of the output into a vectorized representation and reformulate the signal into a coarray format. Subsequently, the digital beamforming (DBF) technique is applied to obtain the image of the coarray spatial spectrum. This transformation of data from a manifold to an image enables the modeling of the problem as an inverse problem, where the objective is to map the spatial spectrum to the sparse receive signal. In order to calibrate the impact of hardware impairments, the resulting spectrum is fed into a convolutional neural network (CNN). Then, by incorporating the sparse constraint into the conjugate gradient algorithm, the spatial spectrum is reconstructed using the proposed sparse conjugate gradient (SCG) algorithm, which provides closed-form solutions. Finally, an iterative optimization process between the CNN and SCG is proposed, resulting in the integration of the model-driven deep learning-based spatial spectrum reconstruction (MoDL-SSR) module, which effectively enhances the performance of spectrum calibration and AoA estimation. The key technical contributions of this work can be summarized as follows.
	
\begin{itemize}
\item We devise a frequency diverse multi-task autoencoder to serve as beamformers across various subcarriers and account for the angular and frequency-dependent hardware impairments. The input channel state information (CSI) is filtered into multiple subregions to augment the underlying statistical consistency within each subregion.
\item We reformulate the spatial spectrum reconstruction problem as a inverse problem, wherein the objective is to map the coarray spectrum to a sparse solution within an over-completed signal set. This formulation leads to the development of the MoDL-SSR module, which incorporates an iterative optimization scheme that involves the interplay between a CNN and the CG algorithm. In particular, the CNN weight coefficients are shared across different iterations to mitigate the computational burden.
\item We modify the  CG algorithm by incorporating a sparse prior, resulting in the proposed SCG algorithm. This modification enables effective recovery of the sparsity of the solution. We further adopt one-hot vectors as labels for training, thus reducing the labeling complexity.
\end{itemize}

The rest of this paper is organized as follows. In Section \ref{sec:SM}, we first describe the ideal signal model under investigation. Then, we introduce the angular-dependent phase errors caused by the hardware impairment and explain the corresponding influence for AoA estimation. In Section \ref{sec:framework}, the overall framework of the proposed MoD-DNN is proposed. We then design a frequency diverse multi-task autoencoder-based beamformer for spatial filtering. The coarray spatial spectrum generation module is also designed to obtain the image of spatial spectrum and formulate the inverse-problem. In Section \ref{sec:MoDLN}, we design the MoDL-SSR module to automatically calibrate the angular-dependent phase error via the iteration between CNN and SCG. In Section \ref{sec:simu}, we evaluate the performance of the proposed scheme via extensive numerical simulations based on 3GPP 5G indoor channel models. Furthermore, we conduct real field tests in an anechoic chamber and an indoor field in Section \ref{sec:experiment}. The conclusion of this paper is drawn in Section \ref{sec:conclusion}.
	
Notations: Lower (upper)-case bold characters are used to denote vectors (matrices), and the vectors are by default in column orientation.  The superscripts $(\cdot)^{\mathsf T}$, and $(\cdot)^{\mathsf H}$ represent the transpose and conjugate transpose operators, respectively. ${\rm {E}}[\cdot]$ denotes the expected value of a discrete random variable. $\lfloor \cdot \rfloor$ denotes the floor function. Symbols $\odot$  stands for Hardmard  product. $\jmath = \sqrt{-1}$ is the imaginary unit.

\section{Preliminary}\label{sec:SM}

\subsection{Direction Finding Based on Ideal Model}
\label{sec:IdealSignal}

We investigate the task of estimating the bearing of mobile terminals using commodity gNodeB by measuring the AoA of emitted electromagnetic signals as they impinge upon the receiving array. Our study considers multiple-input multiple-output orthogonal frequency-division multiplexing (MIMO-OFDM) air interface, as it is the dominant system configuration for 5G broadband wireless communications. We first quantify the field of view (FoV) using a uniform discrete direction set denoted as $\boldsymbol{\theta} = [\theta_1, \ldots, \theta_l, \ldots, \theta_{L}] ^{\mathsf{T}}$, and the angular interval is represented by $\Delta \theta = \theta_l-\theta_{l-1}$. Suppose that the gNodeB employs an $M$-element uniform linear array with a unit inter-element spacing denoted as $d$. Then, the output of the $m$-th sensor and the $k$-th subcarrier is formulated in an over-complete form as
\begin{equation}
\begin{array}{*{20}{l}}
x_{m,k}(t) =\displaystyle\sum\nolimits_{l=1}^{L}\!s_{l,k}(t)\cdot\exp\{{\!-\!\jmath 2\uppi(\! ({f_0\!+\!(m\!\!-\!\!1) \Delta\!f})\!(m\!\!-\!\!1)d \sin\theta_l\!+\!(k\!\!-\!\!1) \Delta\!f r_l)}\!/{{\rm{c}}}\},
\end{array}
\end{equation}
where $\rm{c}$ is the speed of light and $s_{l,k}(t)$ represents the complex waveform of the uplink-sounding reference signal (UL-SRS) impinging upon the system from an AoA $\theta_l$ and a range $r_l$, during a specific time period denoted as $t=1,\ldots,T$. It merits mention that the SRS takes the form of an OFDM signal, distinguished by a pre-determined Zadoff-Chu sequence uniquely designated for each individual terminal \cite{NR2021}. Given the narrow-band system assumption, we have $\Delta f \ll f_0$ and, thus, the noisy receive signal can be further expressed as
\begin{align}	
x_{m,k}(t)\!\approx\!\sum\nolimits_{l=1}^{L}\!s_{l,k}(t){\rm{e}}^{{-\jmath 2\uppi f_0 (m-1)d \sin\theta_l}/{{\rm{c}}}}+n_{m,k}(t),
\end{align}
where $n_{m,k}(t)$ denotes the additive white Gaussian noise (AWGN). Stacking $x_{m,k}(t)$ for all $m=1,\ldots,M$, the receive signal vector of the antenna array for the $k$-th subcarrier is
\begin{align}\label{idealsignal}
\mathbf{x}_{k}(t) \!=\! \sum\nolimits_{l=1}^{L}\mathbf{a}(\theta_l,d)s_{l,k}(t)\!+\!\mathbf{n}(t)\!=\!\mathbf{A}(\boldsymbol{\theta},d)\mathbf{s}_{k}(t)\!+\!\mathbf{n}_{k}(t),
\end{align}
where $\mathbf{a}(\theta_l,d)=[1,\ldots,{\rm{e}}^{{\jmath 2\uppi f_0 (M-1) d \sin\theta_l}/{{\rm{c}}}}]^{\mathsf{T}}$ is the steering vector. We let $\mathbf{A}(\boldsymbol{\theta},d)$$= [$$\mathbf{a}(\theta_1,d),$ $\ldots,$  $\mathbf{a}(\theta_L,d) ]$ denote the array manifold corresponding to the over-complete direction set $\boldsymbol{\theta}$, which is exactly determined by the AoA of the user and the ideal array configuration.  By futher employing the fast Fourier transform (FFT) technique, we are able to obtain the channel frequency response (CFR), or the frequency snapshot, of the $k$-th subcarrier as
\begin{equation}
	\label{eq:hk}
\mathbf{h}(k) = \mathbf{A}(\boldsymbol{\theta},d)\bar{\mathbf{s}}(k)+\bar{\mathbf{n}}(k),\quad  k=1,\ldots,K.
\end{equation}
Here $\bar{\mathbf{s}}(k)$ and $\bar{\mathbf{n}}(k)$ respectively represents the signal and noise in frequency domain. The sample covariance matrix of the received signal is calculated as $\hat{\mathbf{R}}\!\!=\!\! \sum\nolimits_{k\!=\!1}^{K}\!\mathbf{h}(k)\mathbf{h}^{\!\mathsf{H}\!}(k)$, which will be used as input for the estimators. Popular model-driven estimators include digital beamforming (DBF) and multiple signal classification (MUSIC) algorithms, which estimate the AoA respectively as
\begin{align}
\hat{\theta}_{\rm{DBF}} =  \arg\mathop{\max}_{\theta}\mathbf{a}^{\mathsf{H}}(\theta)\hat{\mathbf{R}}\mathbf{a}(\theta) \quad \text{and} \quad  \hat{\theta}_{\rm{MUSIC}} = \arg\mathop{\max}_{\theta} \frac{1}{\mathbf{a}^{\mathsf{H}}(\theta)\mathbf{U}_{\rm{N}}\mathbf{U}^{\mathsf{H}}_{\rm{N}}\mathbf{a}(\theta)}.
\end{align}
The noise subspace ${\mathbf{U}_{\rm{N}}}$ used in MUSIC is constructed using eigenvectors associated with the least eigenvalues of the matrix $\hat{\mathbf{R}}$.

\subsection{Angular-Dependent Phase Errors}

Nevertheless, real-world systems inevitably face hardware impairments, leading to a non-ideal response of the antenna array. As a result, the estimation performance of conventional methods suffers degradation when deployed in practical scenarios. Several types of array error are identified to capture the impact of hardware impairments, primarily including element position errors, gain and phase inconsistencies, and mutual coupling. The cumulative influence of these errors is holistically modelled as
\begin{align}\label{typical_error_model}	\mathbf{h}=\mathbf{C}\mathbf{\Gamma}\mathbf{A}(\boldsymbol{\theta},d+\Delta d)\bar{\mathbf{s}}+\bar{\mathbf{n}},
\end{align}	
where $\Delta d$ represents the element position error, and the mutual coupling coefficients between the array elements are captured by a Toeplitz matrix denoted as $\mathbf{C}$.  Additionally, the gain and phase inconsistencies are incorporated by a matrix $\mathbf{\Gamma}\!\!=\!\!{\rm{diag}}\{\gamma_1,\!\ldots\!,\gamma_L\}$, with $\gamma_l$ being the inconsistency coefficient of the $l$-th RF channel. Under the specific signal model pertaining to hardware impairments, tailored algorithms are developed with the aim of automatically calibrating the phase error within the receive signal.  Under the specific signal model pertaining to hardware impairments, tailored algorithms are developed with the aim of automatically calibrating the phase error within the receive signal. Nevertheless, the array imperfection of a practical system often exceed the scope of these models. In typical scenarios, the hardware impairments induce non-ideal phase errors that exhibit variations across different AoAs.

\begin{figure*}[!htbp]
\centering
\includegraphics[width=1\linewidth]{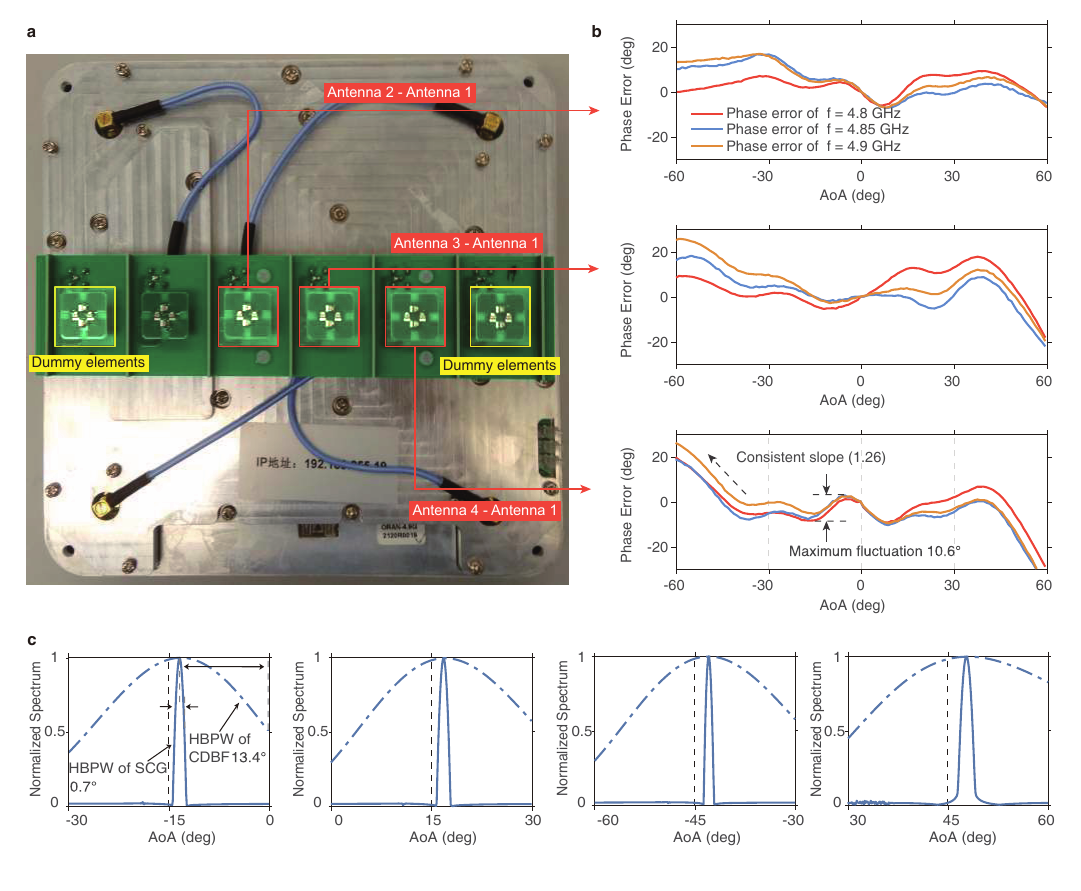}
\caption{Antenna array of a commodity gNodeB for 5G communications. \textbf{a}, We design and fabricate a four-element ULA as RRU antennas. Two antennas at both sides are dummy elements to guarantee the same boundaries seen by the central four elements. \textbf{b}, Measured phase errors of the antennas by selecting the first antenna as reference. Within the AoA ranges spanning from $-30^{\circ}$ to $0^{\circ}$ and from $0^{\circ}$ to $30^{\circ}$, the phase errors demonstrate a relatively stable level of fluctuation, remaining below $20^{\circ}$. However, beyond these subregions, i.e., beyond $-30^{\circ}$ and $30^{\circ}$, the phase errors noticeably escalate while maintaining a consistent slope. \textbf{c}, Comparison of spatial spectrum generated by Coarray DBF (CDBF) and SCG algorithms within the phase errors. Black dashed line denotes the truth. When the AoA is  $-15^{\circ}$, the half-power beam width (HPBW) of CDBF is $13.4^{\circ}$, while that of SCG is $0.7^{\circ}$.}
\label{fig:figureSM1}
\end{figure*}

In order to investigate this phenomenon, we conducted an empirical study by observing the array response of a commodity gNodeB equipped with four antennas, originally designed for 5G communications (see Fig.~\ref{fig:figureSM1}a), within an anechoic chamber setting. By selecting the phase $\phi_1$ of the first antenna as the reference, the phase error $\psi_m$ of the $m$-th antenna can be expressed as $\psi_m(\theta) = \phi_m(\theta)-\phi_1(\theta)-\Delta\phi_{m}(\theta)$, where $\Delta\phi_{m}$ is the ideal phase difference between the $m$-th and the first antennas, which can be determined by computing the wave path difference. As shown in Fig.~\ref{fig:figureSM1}b, we observe that the phase error of the antennas exhibits fluctuations as AoA varies. This observation serves as empirical evidence that the array imperfections in practical systems manifest as angular-dependent phase errors that can exceed $20^\circ$. To precisely incorporate this error into the signal model, we adjust the received signal in CFR form as
\begin{align}\label{angular_dependent_error}
\mathbf{h}(k) &= \sum\nolimits_{l=1}^{L}\boldsymbol{\gamma}_k(\theta_l)\odot\mathbf{a}(\theta_l,d)\bar{s}_l(k)+\bar{\mathbf{n}}(k)=\boldsymbol{\Gamma}_k\odot\mathbf{A}(\boldsymbol{\theta},d)\bar{\mathbf{s}}(k)+\bar{\mathbf{n}}(k),
\end{align}
where $\boldsymbol{\Gamma}_k\!\!=\!\![\boldsymbol{\gamma}_k(\theta_1\!),\!\ldots\!, \boldsymbol{\gamma}_k(\theta_L\!)]$ and $\boldsymbol{\gamma}_k(\theta_l)\!=\! [\gamma_{1,k}(\theta_l),\!\ldots\!,\gamma_{M,k}(\theta_l)]^{\mathsf{T}}$. The coefficient $\gamma_{m,k}(\theta_l)$ represents the phase error of the $m$-th antenna at the $k$-th subcarrier for a given AoA $\theta_l$. Evidently, the typical model \eqref{typical_error_model} can be viewed as a moderate simplification of \eqref{angular_dependent_error}, as it only provides an approximate formulation of the impact caused by practical hardware impairments.

\section{Network Framework and Problem Formulation}
\label{sec:framework}

\begin{figure*}[!htb]
\includegraphics[width=1\linewidth]{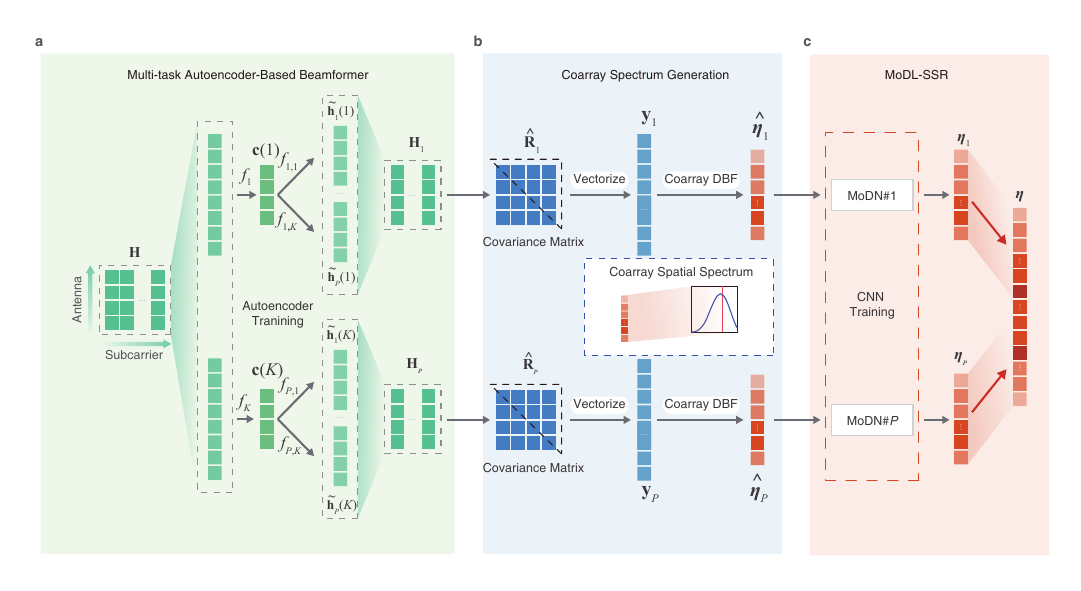}
\caption{Network framework of MoD-DNN for AoA estimation using gNodeB. \textbf{a}, Multi-task autoencoder. We first rearrange the input CSI matrix $\mathbf{H}$ to real-value vectors-form $\mathbf{h}(k)$. Then, we apply a group of multi-task autoencoders to filter the input into $P$ subregions and enhance the consistency of hardware impairments within each subregion.  \textbf{b}, Coarray spectrum generation module.  We vectorize sampled covariance matrix  $\hat{\mathbf{R}}_p$ calculated from the CSI $\mathbf{H}_{p}$ to obtain the coarray signal $\mathbf{y}_p$. The signal is then transformed to the image of the coarray spatial spectrum $\hat{\boldsymbol{\eta}}_p$ by utilizing the coarray DBF technique. \textbf{c}, Model-driven deep learning-based spatial spectrum reconstruction. We integrate deep learning and signal processing methods to design a module with the ability of spectrum reconstruction and enhance the AoA estimation.}
\label{fig:Figure1}
\end{figure*}

Purely model-driven estimators usually struggle to address the challenges posed by a multitude of real-world nonlinear phase errors. In this Article, we propose the model-driven deep neural network (MoD-DNN), a new data-and-model-driven framework for AoA estimation using commodity 5G gNodeB while accounting for the presence of hardware impairments. The proposed framework, as visualized in Fig.~\ref{fig:Figure1},  comprises three cascade-connected modules:  (1) a multi-task autoencoder-based beamformer, (2) a coarray spectrum generation module and (3) a model-driven deep learning-based spatial spectrum reconstruction module.

First, we design a frequency diverse multi-task autoencoder for spatial beamforming. Through encoding and multi-task decoding, the input signal is filtered into different adjacent subregions, thereby enhancing the consistency of hardware impairments within each subregion. In the coarray spectrum generation module, we transform the covariance matrix of the output into a vectorized representation and reformulate the signal into a coarray format. Subsequently, the digital beamforming (DBF) technique is applied to obtain the image of the coarray spatial spectrum. This transformation of data from a manifold to an image enables the modeling of the problem as an inverse problem \cite{MoDL2019}, where the objective is to map the spatial spectrum to the sparse receive signal. In order to calibrate the impact of hardware impairments, the resulting spectrum is fed into a convolutional neural network (CNN). By incorporating the sparse constraint into the conjugate gradient algorithm, the spatial spectrum is reconstructed using the proposed sparse conjugate gradient (SCG) algorithm, which provides closed-form solutions. Finally, an iterative optimization process between the CNN and SCG is proposed, resulting in the integration of the model-driven deep learning-based spatial spectrum reconstruction (MoDL-SSR) module, which effectively enhances the performance of spectrum calibration and AoA estimation. 

In the subsequent subsections, we provide a comprehensive elucidation of each module within the proposed framework.

\subsection{Multi-Task Autoencoder-Based Beamformer}\label{sec:Autoencoder}

As is widely recognized, data consistency holds paramount significance for the efficacy of deep learning methodologies. In the context of the AoA estimation addressed in this study, consistency can be understood as the underlying pattern exhibited by phase errors across various AoAs and frequency points. Drawing upon the measurements conducted within the anechoic chamber, as depicted in Extended Data  Figs.~\ref{fig:Figure1}b--d, the phase error curves distinctly exhibit continuity and regularity within a confined spatial subregion. This empirical evidence reveals that the hardware impairments exhibit similar characteristics across different AoA values within a limited angular range. In light of this finding, we first devise a multi-task autoencoder as beamformers to effectively filter the received signal, aiming to enhance its consistency within the aforementioned small subregion. Compared to the MUSIC-based decision mechanism \cite{Liu20235GNR} which is susceptible to hardware impairment at the boundary of subregions, this approach ensures that the subsequent module receive accurate inputs, which directly impacts the ability. Additionally, as we identify the inconsistency of array imperfections across various frequency points, a frequency diverse autoencoder is further designed for different subcarrriers. The structure of the proposed multi-task autoencoder is illustrated in Fig.~\ref{fig:Figure1}a. The autoencoder first compresses the received signal into a lower-dimensional representation, allowing for the extraction of principal components, followed by the reconstruction of the encoded signal back to its original dimension within $P$ adjacent subregions.

In the network architecture sketched in Fig.~\ref{fig:Figure1}, we consider a single-layer encoder and decoder configuration. In this setup, the $k$-th snapshot of the CFR-form received signal $\mathbf{h}(k)$ first undergoes compression using the fully connected encoder, resulting in
\begin{align}
\mathbf{c}(k) = \tilde{f}_{k}\left[\mathbf{C}_k\tilde{\mathbf{h}}(k)+\mathbf{b}_k\right],
\end{align}
where $\mathbf{C}_k\in\mathbb{R}^{2M\times \vert\mathbf{c}\vert}$ denotes the feedforward computations performed by the encoder for the $k$-th snapshot. $\tilde{\mathbf{h}}(k)\in \mathbb{R}^{2M}$ is the real-value form of the received signal by stacking the real and complex parts of  $\mathbf{h}(k)$. The additive bias vector for the encoding operation is denoted as $\mathbf{b}_k$, and $\tilde{f}_{k}[\cdot]$ represents the elementwise activation function applied to the encoding layer. Following the encoding stage, $P$ decoders are employed to restore $\mathbf{c}_k$ to its original dimension, yielding the output
\begin{align}
\tilde{\mathbf{h}}_{p}(k) = {f}_{p,k}\left[\mathbf{D}_{p,k}{\mathbf{c}}(k)+\mathbf{b}_{p,k}\right],
\end{align}
where $\tilde{\mathbf{h}}_p(k)$ represents the output of the autoencoder specific to the $p$-th subregion. The decoding operation for each subregion involves the feedforward computations $\mathbf{D}_{p,k}$, the elementwise activation function ${f}_{p,k}[\cdot]$, and the additive bias vector $\mathbf{b}_{p,k}$.

To filter the received signal into $P$ subregions, we begin by uniformly partitioning the discrete set of AoA, denoted as $\boldsymbol{\theta} = [\theta_1,\ldots,\theta_L]$, into $P$ subsets as $\boldsymbol{\theta}_1=[\theta_1,\ldots,\theta_{\lfloor L/P \rfloor}]$, $\ldots$ and $\boldsymbol{\theta}_P=[\theta_{\lfloor L(P-1)/P \rfloor+1},\ldots,\theta_L]$. Here $\lfloor \cdot \rfloor$ denotes the floor function. The output of the autoencoder is determined by
\begin{align} \label{autoencoderoutput}
\tilde{\mathbf{h}}_p(k)=
\begin{cases}
\tilde{\mathbf{h}}(k), & \theta \in \boldsymbol{\theta}_p, \\
\quad \boldsymbol{0}, &  {\rm{otherwise}}.
\end{cases}
\end{align}
Following this, the real-value representation $\tilde{\mathbf{h}}_p(k)$ is rearranged to its complex form, resulting in $\mathbf{h}_p(k)$ as the final output of the multi-task autoencoder. Thus far, we have effectively filtered the received signal to partition it into $P$ subregions. Note that the partitioning can be easily extended to a general case based on the varying characteristics of hardware impairments. Then, the output $\mathbf{h}_p(k)$ is restructured into a form known as multiple measurement vector (MMV), denoted as $\mathbf{H}_p=[\mathbf{h}_p(1),\ldots,\mathbf{h}_p(K)]$. In doing so, data consistency within each partition is amplified, which is beneficial to the subsequent network training. As depicted in Fig.~\ref{fig:Figure1}, following this stage, subsequent processing can be carried out independently for each subregion, allowing us to conveniently omit the subscript $p$ indicating the specific subregion from this point onwards.

\subsection{Coarray Spectrum and Inverse Problem Formulation}	

In the preceding subsection, we obtained the received signal $\mathbf{H}$. However, directly extracting features from such representative manifold-form signals, often characterized by high dimensions and intricate topological structures, can be very challenging. In contrast, the spatial spectrum, presented in the form of an image, holds the capability to decipher the attributes of hardware impairments from diverse facets such as beamwidth, kurtosis, and skewness. Hence, by converting the received signal into spatial spectrum, we gain the advantage of tapping into the inherent structure and underlying correlations intrinsic to $\mathbf{H}$, thereby providing a more revealing insight into the signal features.

Considering the discrete AoA set $\boldsymbol{\theta}$, the covariance matrix of the received signal can be represented as
\begin{align}
\mathbf{R}={\rm{E}}\left[\mathbf{h}(k)\mathbf{h}^{\mathsf{H}}(k)\right] = \sum\nolimits_{l=1}^{L}\eta_l\mathbf{a}(\theta_l)\mathbf{a}^{\mathsf{H}}(\theta_l)+\sigma^2_{\rm{n}}\mathbf{I},
\end{align}
where $\eta_l$ signifies the signal power for direction $\theta_l$, $\sigma_n^2$ represents the power of the AWGN, and $\mathbf{I}$ denotes the identity matrix. We further rewrite the $m$-th column of $\mathbf{R}$ as
\begin{align}
\mathbf{r}_m= \sum\nolimits_{l=1}^L\eta_l\mathbf{a}(\theta_l)\mathbf{a}^{\mathsf{H}}(\theta_l)\mathbf{e}_m+\sigma^2_{\rm{n}}\mathbf{I}\mathbf{e}_m=\mathbf{A}_m\boldsymbol{\eta}+\sigma^2_{\rm{n}}\mathbf{e}_m,
\end{align}
where $\mathbf{A}_m = [\mathbf{a}(\theta_1)\mathbf{a}^{\mathsf{H}}(\theta_1)\mathbf{e}_m,\ldots,\mathbf{a}(\theta_L)\mathbf{a}^{\mathsf{H}}(\theta_L)\mathbf{e}_m]$, $\boldsymbol{\eta}=[\eta_1,\ldots,\eta_L]$ can be interpreted as the sparse spatial spectrum. The vector $\mathbf{e}_m\in\mathbb{R}^{M}$is constructed such that its $m$-th element is 1 while all other elements are zero. Vectorizing the covariance matrix by stacking all the columns of $\mathbf{R}$, we obtain the corresponding coarray signal
\begin{align}
\mathbf{y}&=\left[\mathbf{r}_1;\ldots;\mathbf{r}_M\right]=\left[\mathbf{A}_1;\ldots;\mathbf{A}_M\right]\boldsymbol{\eta}+\sigma_{\rm{n}}^2\left[\mathbf{e}_1;\ldots;\mathbf{e}_M\right]=\tilde{\mathbf{A}}\boldsymbol{\eta}+\sigma^2_{\rm{n}}\tilde{\mathbf{e}},
\end{align}
where $\tilde{\mathbf{A}} = \left[\mathbf{A}_1;\ldots;\mathbf{A}_M\right]$ represents the equivalent coarray manifold. In general, the spatial spectrum can be estimated using DBF techniques, resulting in the estimation
\begin{align}
\label{coarrayspectrum}
\hat{\boldsymbol{\eta}} = \tilde{\mathbf{A}}^{\mathsf{H}}\mathbf{y}\approx \tilde{\mathbf{A}}^{\mathsf{H}}\tilde{\mathbf{A}}\boldsymbol{\eta}=\mathbf{P}\boldsymbol{\eta}.
\end{align}
It is noteworthy that the above equation is also the normal equation form of $\mathbf{y}=\tilde{\mathbf{A}}\boldsymbol{\eta}$. In this context, the original problem can be equivalently stated as the task of solving for $\boldsymbol{\eta}$ given the observed spectrum $\hat{\boldsymbol{\eta}}$, with the \emph{a priori} projection matrix ${\mathbf{P}} = \tilde{\mathbf{A}}^{\mathsf{H}}{\tilde{\mathbf{A}}}$.

The conversion of the spatio-temporal series data into spatial spectrum representation aligns them more closely with conventional image data. This shift has noteworthy implications: it facilitates the utilization of established computer vision techniques, such as CNN, for effective error calibration, capitalizing on their strengths of handling structured data.

\section{MoDL-SSR-Based Spectrum Calibration and AoA Estimation}
\label{sec:MoDLN}

\begin{figure*}[!ht]
\centerline{\includegraphics[width=1\linewidth]{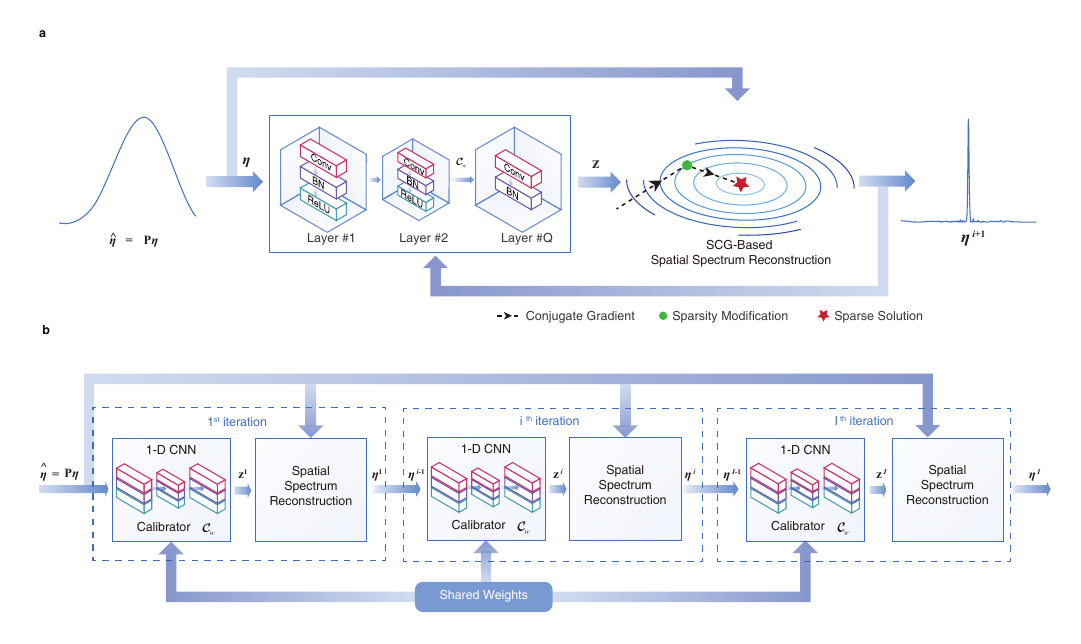}}
\caption{Illustration of proposed iteration for MoDL-SSR module. \textbf{a}, Overall iteration in MoDL-SSR module.  We leverage 1D-CNN to calibrate the input coarray spectrum and obtain a calibrated version $\mathbf{z}$. We then design a SCG algorithm with sparsity modification in CG method. Coarray spectrum is fed into iterations between 1D-CNN and SCG algorithm to reconstruct sparse spatial spectrum.  \textbf{b}, Alternative optimization between CNN-based calibrator and inverse problem-based SSR. We apply the same 1D-CNN at each iteration to share the weights among the CNNs in different iterations.}
\label{fig:Figure2}
\end{figure*}

This section is dedicated to the construction of a specialized neural network aimed at rectify the impact of angular-dependent phase errors within $\mathbf{H}$. To handle the image-form spectrum generated in the previous subsection, we implement and train a CNN tailored for each subregion to effectively calibrate the phase errors present in the coarray signal. However, obtaining accurate spatial spectrums as the ground truth for network training presents a notable labeling complexity. To address this challenge, a modified SCG algorithm is devised to reconstruct the sparse spatial spectrum. One-hot vectors are then employed as surrogate labels. In addition, an iterative feedback mechanism is established between the CNN and SCG, forming the MoDL-SSR framework. This framework enables the continuous refinement of the calibration process through the feedback of SCG outputs into the CNN. The proposed MoDL-SSR framework as a whole harnesses the synergistic interplay between these components to yield improved accuracy in estimating the sparse spatial spectrum.

\subsection{Alternative Optimization Formulation}

To estimate the AoA from the observed coarray signal affected by hardware impairments, we propose a model-driven deep neural network that incorporates iterations between deep learning and signal processing methods. The overall iteration is depicted in Fig.~\ref{fig:Figure2}a. We start by recasting the reconstruction of the spatial spectrum as the following optimization problem
\begin{align}\label{optimization}
\boldsymbol{\eta} = \mathop{\arg\min_{\mathbf{\boldsymbol{\eta}}}}\|\mathbf{P}\boldsymbol{\eta}-\hat{\boldsymbol{\eta}}\|_2^2+\lambda\|{\mathcal{E}}_{\mathbf{w}}(\boldsymbol{\eta})\|^2,
\end{align}
where $\mathcal{E}_{\mathbf{w}}(\boldsymbol{\eta})$ represents a regularization term that accounts for the combined effects of noise and hardware impairments, and it can be learned using a $\mathbf{w}$-weighted CNN. The regularization term can be rewritten as
\begin{align}\label{regularizationterm}
\mathcal{E}_{\mathbf{w}}(\boldsymbol{\eta}) = (\mathbf{I}-\mathcal{C}_{\mathbf{w}})(\boldsymbol{\eta}) = \boldsymbol{\eta} - \mathcal{C}_\mathbf{w}(\boldsymbol{\eta}),
\end{align}
where $\mathcal{C}_\mathbf{w}(\boldsymbol{\eta})$ corresponds to a calibrated version of $\boldsymbol{\eta}$ after mitigating the impact of hardware impairments and noise. By substituting \eqref{regularizationterm} into \eqref{optimization}, we obtain the optimization problem in the $i$-th iteration:
\begin{align}\label{optimization2}
\boldsymbol{\eta}^{i+1} = \mathop{\arg\min_{\mathbf{\boldsymbol{\eta}}}}\|\mathbf{P}\boldsymbol{\eta}-{\boldsymbol{\eta}}^{i}\|_2^2+\lambda\|\boldsymbol{\eta} - \mathcal{C}_{\mathbf{w}}(\boldsymbol{\eta})\|^2.
\end{align}
The calibration term $\mathcal{C}_{\mathbf{w}}(\boldsymbol{\eta}^i+\Delta \boldsymbol{\eta})$ can be further approximated by employing the Taylor series expansion, i.e.,
\begin{align}
\mathcal{C}_\mathbf{w}(\boldsymbol{\eta}^i+\Delta \boldsymbol{\eta}) = \mathcal{C}_\mathbf{w}(\boldsymbol{\eta}^i)+\mathbf{J}_i^{\mathsf{T}}\nabla\boldsymbol{\eta},
\end{align}
where $\mathbf{J}$ is the Jacobian matrix. Upon introducing the updated variable $\boldsymbol{\eta} = \boldsymbol{\eta}^i+\Delta \boldsymbol{\eta}$, the regularization term is further expressed as
\begin{align}	\|\boldsymbol{\eta}-\mathcal{C}_\mathbf{w}(\boldsymbol{\eta}^i+\Delta\boldsymbol{\eta})\|^2 \overset{\Delta\boldsymbol{\eta}\rightarrow\boldsymbol 0}\approx \|\boldsymbol{\eta}-\mathcal{C}_\mathbf{w}(\boldsymbol{\eta}^i)\|^2 + \|\mathbf{J}_i\Delta \boldsymbol{\eta}\|^2.
\end{align}
Then, the optimization problem \eqref{optimization2} can be transformed into the following equivalent:
\begin{subequations}\label{alternativeoptimization}
	\begin{equation}\label{iteration1}
	\mathbf{z}^i=\mathcal{C}_{\mathbf{w}}(\boldsymbol{\eta}^i),
	\end{equation}
	\begin{equation}\label{iteration2}
	\boldsymbol{\eta}^{i+1} = \mathop{\arg\min_{\mathbf{\boldsymbol{\eta}}}}\|\mathbf{P}\boldsymbol{\eta}-{\boldsymbol{\eta}}^{i}\|_2^2+\lambda\|\boldsymbol{\eta} - \mathbf{z}^i\|^2.
	\end{equation}
\end{subequations}
The iterative process is presented in Fig.~\ref{fig:Figure2}b. By examining the alternative optimization in \eqref{alternativeoptimization}, we highlight the following remarks.

$\textit{Remark 1}$: The proposed MoDL-SSR contains two parts: (i) a CNN-based spectrum calibrator, which leverages deep learning methodologies, and (ii) an inverse problem-based spatial spectrum reconstruction, which can be effectively solved using model-based signal processing algorithm. Within this process, the calibrated version $\mathbf{z}^i$ is derived by subjecting $\boldsymbol{\eta}^i$ to the CNN $\mathcal{C}_{\mathbf{w}}$ in \eqref{iteration1}, subsequently serving as the reference for reconstruction in equation \eqref{iteration2}. Recall that the projection matrix $\mathbf{P}$ is generated from the ideal array manifold $\mathbf{A}$, as described in \eqref{eq:hk}. As such, the underlying array manifold of $\boldsymbol{\eta}$ is progressively calibrated towards the ideal manifold to mitigate the impact of hardware impairments.

$\textit{Remark 2}$: The regularization term $\|{\mathcal{E}}_{\mathbf{w}}(\boldsymbol{\eta})\|^2$ assumes larger values when the received signal is contaminated by hardware impairments. To strike a balance between calibration and reconstruction, the regularization coefficient $\lambda$ can be designated as a trainable parameter. However, it is important to emphasize that the proposed framework demonstrates robustness to the choice of $\lambda$, particularly when the hardware impairments are perfectly calibrated as $\boldsymbol{\eta}\!\!\rightarrow\!\!\mathbf{z}$. As a result, the regularization coefficient can also remain fixed throughout the iterations, yielding consistent outcomes.

\subsection{SCG-Based SSR Layer}

Next, we elaborate the SCG algorithm used in the SSR layer. The goal is to reconstruct the spatial spectrum $\boldsymbol{\eta}$ by solving the sub-problem \eqref{iteration2} through the use of the following normal equation
\begin{align}\label{normalequation}
(\mathbf{P}+\lambda\mathbf{I})\boldsymbol{\eta} = \boldsymbol{\eta}^{i}+\lambda\mathbf{z}^{i}.
\end{align}
The CG algorithm is commonly employed in forward models of this type to efficiently tackle such problems. In the following, we first provide a brief overview of the CG algorithm. Specifically, in each iteration of the CG algorithm, a $\mathbf{P}$-conjugate direction $\mathbf{c}(n)$ is selected from the negative of the gradient vector $\mathbf{g}(n)$. The solution is then updated along the direction $\mathbf{c}(n)$ with a step size denoted as $\alpha(n)$. Concretely, the CG steps are given as follows
\begin{subequations}
	\begin{equation}
	{\alpha}(n) = -\frac{\mathbf{g}^{\mathsf{T}}(n)\mathbf{c}(n)}{\mathbf{c}^{\mathsf{T}}(n)\mathbf{P}\mathbf{c}(n)},
	\end{equation}
	\begin{equation}
	\boldsymbol{\eta}(n+1) = \boldsymbol{\eta}(n)+{\alpha}(n)\mathbf{c}(n),
	\end{equation}
	\begin{equation}	{\beta}(n)=\frac{(\mathbf{g}(n+1)-\mathbf{g}(n))^{\mathsf{T}}\mathbf{g}(n+1)}{\mathbf{g^{\mathsf{T}}}(n)\mathbf{g}(n)},
	\end{equation}
	\begin{equation}
	\mathbf{c}(n+1)=-\mathbf{g}(n+1)+{\beta}(n)\mathbf{c}(n).
	\end{equation}
\end{subequations}
We readily see that all steps of the CG algorithm have closed-form solutions, which removes the need for additional parameter training. Moreover, the gradients can be backpropagated from the CG sub-blocks during the iterations, enabling end-to-end training. By leveraging the CG algorithm, the MoDL-SSR modules feature sub-blocks that include a numerical optimization layer with a low computational cost.

While the CG algorithm efficiently solves the problem, we must acknowledge that the original form of the normal equation is given by $\mathbf{y} = \tilde{\mathbf{A}}\boldsymbol{\eta}$, where $\tilde{\mathbf{A}}$ represents an over-complete coarray manifold associated with the discrete AoA set $\boldsymbol{\theta}$. This implies that the inverse problem is inherently underdetermined, leading to a wide beam-width for the solution when using the CG algorithm alone. In this case, the utilization of one-hot vectors as training labels remains impractical. To counteract this, we leverage the inherent sparsity of the solution to narrow the beam-width. To this end, we introduce a sparsity-constrained CG algorithm tailored for spectrum reconstruction. In order to fully utilize the sparsity of the incident signal, we introduce an additional regularization term to the optimization problem \eqref{iteration2} in the form of a sparsity function $s(\boldsymbol{\eta})$ as
\begin{align}\label{optimization_scg}
\boldsymbol{\eta}^{i+1} = \mathop{\arg\min_{\mathbf{\boldsymbol{\eta}}}}\|\mathbf{P}\boldsymbol{\eta}-{\boldsymbol{\eta}}^{i}\|_2^2+\lambda\|\boldsymbol{\eta} - \mathbf{z}^i\|^2+\mu s(\boldsymbol{\eta}),
\end{align}
where $\mu$ denotes the regularization coefficient to enforce the sparsity constraint. The objective is then to minimize this problem through an iterative method that alternates between CG solutions and sparsity modification. The first two terms in \eqref{optimization_scg} are minimized using CG method to obtain an intermediate solution
\begin{align}
\tilde{\boldsymbol{\eta}}(n+1) = \boldsymbol{\eta}(n)+{\alpha}(n)\mathbf{c}(n).
\end{align}
Then, the solution $\tilde{\boldsymbol{\eta}}(n+1)$ is modified using proximal gradient search as
\begin{align}
\boldsymbol{\eta} = \mathop{\arg\min_{\mathbf{\boldsymbol{\eta}}}} \|\boldsymbol{\eta}-\tilde{\boldsymbol{\eta}}(n+1)\|^{2}+\mu s(\boldsymbol{\eta}).
\end{align}
As such, the solution is obtained as
\begin{align}
\boldsymbol{\eta}(n+1) = \tilde{\boldsymbol{\eta}}(n+1) - \mu \nabla^{s}s(\tilde{\boldsymbol{\eta}}(n+1)),
\end{align}
where $\nabla^{\mathrm{s}}s(\tilde{\boldsymbol{\eta}}(n+1))$ is the subgradient of $s(\boldsymbol{\eta})$ at $\tilde{\boldsymbol{\eta}}(n+1)$. Further, we assume that the subgradient $\nabla^{\mathrm{s}}s(\tilde{\boldsymbol{\eta}})$ gradually evolves during the iterations to obtain the approximation  $\nabla^{\mathrm{s}}s(\tilde{\boldsymbol{\eta}}(n+1)) \approx \nabla^{\mathrm{s}}s(\tilde{\boldsymbol{\eta}}(n))$. As such, the two-step iteration can be simplified as
\begin{align}
\tilde{\boldsymbol{\eta}}(n+1) = \boldsymbol{\eta}(n)+{\alpha}(n)\mathbf{c}(n) - \nabla^{\mathrm{s}}s(\tilde{\boldsymbol{\eta}}(n)).
\end{align}
Without loss of generality, we employ a reweighted zero attracting function as
\begin{align}
s(\boldsymbol{\eta}(n)) = {\rm{log}}\left(1+\frac{\|\boldsymbol{\eta}(n)\|_1}{\epsilon}\right),
\end{align}
where $\epsilon$ is an approximation parameter. The corresponding subgradient function in expressed as
\begin{align}
\nabla^{s}s(\tilde{\boldsymbol{\eta}}(n)) = \frac{{\rm{sgn}}(\boldsymbol{\eta}(n))}{1+\epsilon\|\boldsymbol{\eta}(n)\|_1},
\end{align}
where $\rm{sgn}(\cdot)$ is the signum function. We summarize the algorithm in {\bf Algorithm \ref{alg_SCG}}.
	
\begin{algorithm}[!ht]
\caption{SCG-based SSR algorithm.}
\label{alg_SCG}
\begin{flushleft}
\textbf{Input:} Coarray spectrum ${\boldsymbol{\eta}}^{i}$, calibrated spectrum $\mathbf{z}^{i}$, projection matrix $\mathbf{P}$, $\lambda$, $\epsilon$.\\
\textbf{Output:} Reconstructed spatial spectrum $\boldsymbol{\eta}^{i+1}$.\\
\textbf{Initialize:} $\boldsymbol{\eta}^{i+1}(0)=\mathbf{0}$, $\mathbf{g}(0) = (\mathbf{P}+\lambda\mathbf{I})\boldsymbol{\eta}^{i+1}(0)-({\boldsymbol{\eta}}^{i}+\lambda\mathbf{z}^{i})$, $\mathbf{c}(0)=-\mathbf{g}(0)$, $\gamma_{\rm{CG}}$ and $N^{\rm{CGiter}}_{\rm{max}}$.
\end{flushleft}
\begin{algorithmic}[1]
\For{$n=0$ to $N^{\text{iter}}_{\rm{max}}$}\\
${\alpha}(n) = -\frac{\mathbf{g}^{\mathsf T}(n)\mathbf{c}(n)}{\mathbf{c}^{\mathsf T}(n)\mathbf{P}\mathbf{c}(n)}$\\
$\boldsymbol{\eta}^{i+1}(n+1) = \boldsymbol{\eta}^{i+1}(n)+{\alpha}(n)\mathbf{c}(n)-\frac{{\rm{sgn}}(\boldsymbol{\eta}^{i+1}(n))}{1+\epsilon\|\boldsymbol{\eta}^{i+1}(n)\|_1}$\\
$\mathbf{g}(n+1) = (\mathbf{P}+\lambda\mathbf{I})\boldsymbol{\eta}^{i+1}(n)-({\boldsymbol{\eta}}^{i}+\lambda\mathbf{z}^{i})$\\	${\beta}(n)=\frac{(\mathbf{g}(n+1)-\mathbf{g}(n))^{\mathsf T}\mathbf{g}(n+1)}{\mathbf{g^{\rm{T}}}(n)\mathbf{g}(n)}$\\
			$\mathbf{c}(n+1)=-\mathbf{g}(n+1)+{\beta}(n)\mathbf{c}(n)$
\If{					$\boldsymbol{\eta}^{i+1}(n+1)-\boldsymbol{\eta}^{i+1}(n)<\gamma_{\rm{CG}}$}\\
			\qquad \quad Break;
\EndIf
\EndFor
\end{algorithmic}
\end{algorithm}

\subsection{Training of CNN-based Calibrator}

	In the previous subsection, we introduced the SCG algorithm, capable of recovering the sparse AoA solution from an ideal spatial spectrum. The final and crucial phase in achieving a closed-loop process involves the design of a neural network to calibrate the spatial spectrum with phase errors. In this subsection, we present the architecture of the CNN used as the spectrum calibrator in \eqref{iteration1}. As shown in Fig.~\ref{fig:Figure2}a, the spatial spectrum $\boldsymbol{\eta}^{i}$ is passed to the CNN input layer. The CNN consists of $P$ layers, each layer comprising a convolution operation followed by batch normalization and a rectified linear unit (ReLU) activation function. The final layer does not employ ReLU activation, allowing the network to learn the negative part of the array imperfection patterns. In this network framework, a four-layer CNN structure is employed, with each dimension being $32 \times 1$, and the number of convolutional kernels being $4$, $8$, $4$, and $1$, respectively. Concretely, the output of the $p$-th layer is expressed as
\begin{align} \label{CNNoutput}
\mathbf{c}=
\begin{cases}
{\rm{ReLU}}\left(\mathbf{w}_p*\mathcal{P}(\mathbf{c}_{p-1})\right), & p = 1,\ldots,3, \\
\mathbf{w}_p*\mathcal{P}(\mathbf{c}_{p-1}), &  p = 4.
\end{cases}
\end{align}

The first input is $\mathbf{c}_{0}=\boldsymbol{\eta}^{i-1}$, and the calibrated spectrum is obtained as  $\mathbf{z}^{i} = \mathbf{c}_P$. The parameter $\mathbf{w}_p$ denotes the convolution kernel corresponding to the $p$-th layer. The input $\mathbf{c}_{p-1}$ is zero-padded at the borders using the padding operator $\mathcal{P}(\cdot)$ to implement the convolution operation. Then, $\mathbf{z}^i$ is fed into the SCG algorithm to yield the sparse solution $\boldsymbol{\eta}^{i}$. Upon completion of the iterations between the CNN and SCG modules, the reconstructed spectrum $\boldsymbol{\eta}^{I}$ is considered as the output of the framework.

During network training, the objective is to minimize the mean square error (MSE) loss function between the coarray signal $\mathbf{y}$ and the one-hot vector $\boldsymbol{\eta}$ over the training dataset $\mathbb{D}_{\text{train}} = \left\{(\mathbf{y}(1),\boldsymbol{\eta}(1)),\ldots,(\mathbf{y}(D),\boldsymbol{\eta}(D)) \right\}$, i.e.,
\begin{align}
\mathcal{L} = \sum\nolimits_{d=1}^{D}\|\boldsymbol{\eta}^{I}(d)-\boldsymbol{\eta}(d)\|^2,
\end{align}
where $\boldsymbol{\eta}^{I}$ denotes the spectrum recovered from $\mathbf{y}$ using the reconstruction framework described earlier. To determine the parameters $\mathbf{w}$ of the CNN-based calibrator, the gradient of the loss function with respect to $\mathbf{w}$ can be computed using the chain rule as
\begin{align}
\nabla_\mathbf{w}\mathcal{L} = \sum\nolimits_{i=1}^{I}\mathbf{J}^{\mathsf{T}}_{\mathbf{w}}(\mathbf{z}^{i-1})\mathbf{J}^{\mathsf{T}}_{\mathbf{z}^{i-1}}(\boldsymbol{\eta}^{i})(\nabla_{\boldsymbol{\eta}^i}\mathcal{L}).
\end{align}
We derive from \eqref{normalequation} that $\boldsymbol{\eta}^{i} = (\mathbf{P}+\lambda\mathbf{I})^{-1}(\mathbf{z}^{i-1}+\lambda\boldsymbol{\eta}^{i-1})$. The above calculations rely on the convergence of the SCG algorithm, allowing for backpropagation of gradients through the SCG block. Notice that, the same CNN-based calibrator $\mathcal{C}_{\mathbf{w}}$ is used at each iteration, meaning that the weights are shared among the CNNs in different iterations. This shared parameterization ensures consistency and facilitates efficient training of the CNN-based calibrator.

\section{Performance Evaluation via Simulations}
\label{sec:simu}

We evaluate the performance of the proposed MoD-DNN method for AoA estimation in the presence of hardware impairments. The CSI data used in the numerical studies is generated using a state-of-the-art link-level wireless communications simulator \cite{Jia2022Link}. The simulator is capable of accommodating customized hardware impairment functions and can replicate channels aligned with the LoS-only scenario as stipulated in the 3GPP TR 38.901 standard \cite{Study3GPP2020}. Our focus centers on a typical four-antenna gNodeB configuration, which mirrors the most common setup in current commercial 5G deployments in indoor environment. The system and waveform parameters used in the simulations are detailed in Table~\ref{table1}. We consistently employ the indoor factory LoS channel at a Sub-6 GHz working frequency (denoted as 3GPP\_38.901\_InF\_LoS) across all simulations.

\begin{table}[htbp!]
	\centering
	\caption{Settings of 5G System and UL-SRS}
	\label{table1}
	\begin{tabular}{c c c c c}
    \toprule
		& & &\\[-10pt]
		\textbf{Parameter}&\textbf{Value}&\textbf{Parameter}&\textbf{Value}\\
		\hline
		&\\[-6pt]
		Type of gNB& Picocell gNB&Number of subcarriers&3264\\
		&\\[-6pt]
		Number of antenna elements&4&Bandwidth&$100$ MHz\\	
		&\\[-6pt]
		Array type&ULA&UL-SRS pattern&Comb-two\cite{NR2021}\\
		&\\[-6pt]
		Central frequency&$4.85$ GHz&Sampling frequency&$122.88$ MHz\\
		&\\[-6pt]
		Subcarrier spacing&$30$ kHz&UL-SRS temporal interval&$80$ ms\\
  \bottomrule
	\end{tabular}
\end{table}

To mitigate storage constraints, we uniformly sample $16$ subcarriers from the CSI dataset. Illustrations of the angular-dependent phase error for this configuration can be found in Fig.~\ref{fig:figureSM1}b, while the comprehensive data for each subcarrier is accessible at \cite{Pan2022Data}. The AoA of the user equipment (UE) varies within the range of $\left[-60^{\circ}, 60^{\circ}\right]$, with a uniform interval $\Delta \theta = 0.1^{\circ}$. Each AoA value is associated with $60$ time slots, resulting in a total of $1201\times50 = 60050$ groups of CSI data for analysis.  During the network training using PyTorch, we employ the Adam optimization scheme to iteratively update network parameters, which allows the computation of the weight gradients using backpropagation \cite{Kingma2015Adam}. The learning rate and mini-batch size are set as $0.01$ and $64$, respectively. Subsequently, the learning rate undergoes reduction by a factor of $0.5$ every $5$ epochs. Empirical choices for the regularization coefficient $\lambda$ and the weight coefficient $\epsilon$ are $0.1$ and $0.5$, respectively. The number of epochs is set at $30$, except for the simulation in Fig.~\ref{fig:figureSM2}h, where the number of epochs varies.

\subsection{Performance of Multi-task Autoencoder}

\begin{figure*}[!ht]
\centerline{\includegraphics[width=1\linewidth]{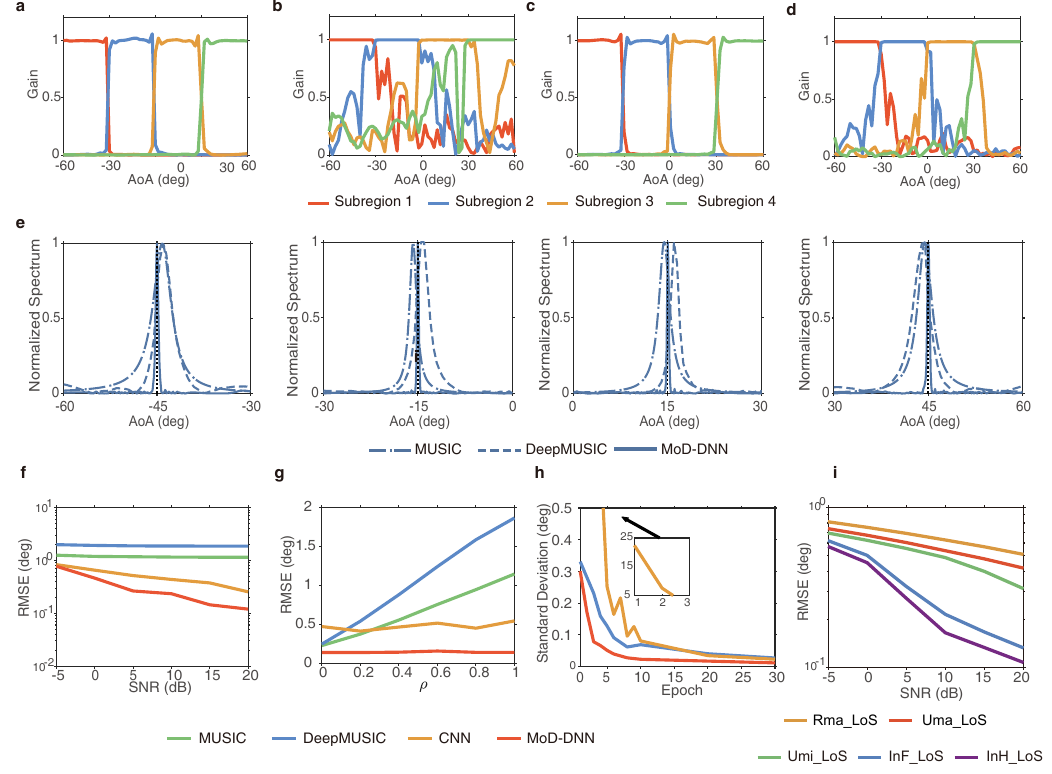}}
\caption{Performance comparison based on dataset for numerical experiments. \textbf{a},  Amplitude responses of Autoencoder \uppercase\expandafter{\romannumeral1}. \textbf{b},  Phase responses of Autoencoder \uppercase\expandafter{\romannumeral1}. \textbf{c},  Amplitude responses of Autoencoder \uppercase\expandafter{\romannumeral2}. \textbf{d},  Phase responses of Autoencoder \uppercase\expandafter{\romannumeral2}. \textbf{e}, Comparison of spatial spectrum results at different AoA. Black dashed lines denote the truth. \textbf{f}--\textbf{h}, Performance comparison between different methods. \textbf{f}, RMSE versus SNR. \textbf{g}, RMSE performance comparison versus degree of impairment $\rho$. \textbf{h}, SD performance comparison versus number of epoch. \textbf{i}, RMSE performance of MoD-DNN in different channels.}
\label{fig:figureSM2}
\end{figure*}

\begin{table}[htbp!]
\centering
\caption{Settings of Multitask Autoencoder}
\label{table2}
\begin{tabular}{cccc}
\toprule
	& & & \\[-8pt]
	Training Set&$[-60^{\circ}\!:\!0.1^{\circ}\!:\!60^{\circ}]$&Validation set&$[-60^{\circ}\!:\!0.1^{\circ}\!:\!60^{\circ}]$\\
	\hline
	& & & \\[-6pt]
	Size&$1201 \times 40$&Size&$1201 \times 10$\\
	\hline
	& & & \\[-6pt]
	Layer 1&$16\times16\times8$& Layer 2& $16\times8\times16$\\
\bottomrule
\end{tabular}
\end{table}

In the first set of tests, we evaluate the beamforming performance achieved by the proposed frequency diverse multi-task autoencoder. As a comparison, we also include a multi-task autoencoder that directly processes the original CSI across all frequency points, disregarding the phase differences between subcarriers. For convenience, we refer to these as Autoencoder \uppercase\expandafter{\romannumeral1} and  Autoencoder \uppercase\expandafter{\romannumeral2}, respectively. We collect $1201\times40 = 48040$ groups of simulated CSI data for training, and an additional $1201\times10 = 12010$ groups for validation, with a fixed input signal-to-noise ratio (SNR) of $10\;\text{dB}$. To substantiate the effectiveness of the proposed spatial filtering method while conserving computational resources, the spatial scope of $[-60^{\circ},60^{\circ}]$ is evenly divided into $P=4$ subregions, i.e., $[-60^{\circ},-30^{\circ})$, $[-30^{\circ},0^{\circ})$, $[0^{\circ},30^{\circ})$ and $[30^{\circ},60^{\circ}]$. The detailed autoencoder settings can be found in TABLE~\ref{table2}. Note that, the number of the subregions can be conveniently adjusted according to different requirements. The spatial responses of the multi-task autoencoder are provided in Figs.~\ref{fig:figureSM2}a--d. The amplitude and phase responses of the multi-task autoencoder-based beamformers are defined as
\begin{equation}
r_\text{amp}^{p} = |\mathbf{h}^{\mathsf H}(k)\mathbf{h}_p(k)| \quad \text{and} \quad r_\text{phase}^{p} = \frac{|\mathbf{h}^{\mathsf H}(k)\mathbf{h}_p(k)|}{\|\mathbf{h}(k)\|_2\|\mathbf{h}_p(k)\|_2}.
\end{equation}

Figs.~\ref{fig:figureSM2}a and c show the amplitude responses of Autoencoder \uppercase\expandafter{\romannumeral1} and  Autoencoder \uppercase\expandafter{\romannumeral2}, respectively. It is evident that both Autoencoder \uppercase\expandafter{\romannumeral1} and Autoencoder \uppercase\expandafter{\romannumeral2} demonstrate a high degree of agreement in their amplitude responses with the anticipated output across the segmented subregions. Furthermore, their responses exhibit rapid attenuation at the subregion boundaries. In terms of phase response, as depicted in Figs.~\ref{fig:figureSM2}b and d, it becomes apparent that Autoencoder \uppercase\expandafter{\romannumeral1} shows a clearer inconsistency with the input beyond the designated subregion, which indicates a performance superiority compared to Autoencoder \uppercase\expandafter{\romannumeral2}. This superiority can be attributed primarily to the frequency diverse scheme, which effectively mitigates phase errors that fluctuate with subcarriers. The above results collectively underscore the effectiveness of the proposed multi-task autoencoder in segregating signals into distinct subregions to enhance underlying data consistency, thus enabling subsequent signal processing and training procedures to be conducted independently for each subregion.

\subsection{Performance of Spectrum Calibrator}
	
In this subsection, we first present the spectrums generated by coarray DBF and SCG algorithm, which are respectively the input and output of MoDL-SSR module.  For a UE AoA of $-15^{\circ}$, Fig.~\ref{fig:figureSM1}c indicates that the azimuth of UE can be directly estimated from the coarray spectrum. However, the accuracy deteriorates evidently in the presence of hardware impairment. Then, we assess the effectiveness of the proposed SCG algorithm. As demonstrated in Fig.~\ref{fig:figureSM1}c, the application of the sparsity constraint leads to a narrower spectrum width. Nonetheless, the obtained result is not entirely sparse due to the mismatch between the real and ideal coarray manifolds. Similar outcomes are observed in Fig.~\ref{fig:figureSM1}c, when the AoA of the UE is $-45^{\circ}$, $15^{\circ}$ and $45^{\circ}$. This reaffirms the capability of the SCG algorithm to achieve reasonably accurate results despite the presence of hardware impairment. If we calibrate the coarray spectrum beforehand, the SCG algorithm is promising to recover the sparse spatial spectrum and enhance AoA estimation performance. Therefore, by integrating the SCG algorithm, we can employ the corresponding one-hot vectors as training labels, marking a substantial reduction in labeling complexity.

Next, we proceed to assess the efficacy of the proposed method for mitigating impairments by comparing the spatial spectra obtained using different algorithms. Specifically, we consider a UE positioned at $-45 ^{\circ}$. Three different approaches, namely MUSIC, DeepMUSIC, and MoD-DNN, are compared for UE localization. The estimation outcomes are presented in Fig.~\ref{fig:figureSM2}e. As a representative model-driven approach, the MUSIC algorithm merely offers a coarse estimate of the UE's AoA. Furthermore, as a super-resolution algorithm, the MUSIC algorithm is susceptible to model mismatch, leading to broadened spatial spectrum beams. This is a typical behavior of the algorithm being affected by phase errors. Such phenomenon arises from the simplistic calculation of the predetermined steering vector $\mathbf{a}(\theta)$ based on idealized antenna spacing, which does not strictly exhibit orthogonality with the derived noise subspace $\mathbf{U}_{\mathrm{N}}$. Akin to the MUSIC algorithm, DeepMUSIC's performance is curtailed without adequate phase error calibration. Meanwhile, the spatial spectrum of the DeepMUSIC algorithm displays increased fluctuations compared to the original MUSIC algorithm, indicating heightened vulnerability to phase errors. In contrast, the proposed MoD-DNN excels in accurately reconstructing the spectrum and obtaining nearly unbiased estimations. Furthermore, the yielding spectrum exhibits exact sparsity compared to the SCG algorithm's outcome shown in Fig.~\ref{fig:figureSM2}e. This phenomenon reveals that the input spectrum perfectly matches the corresponding projection matrix in the SCG algorithm, implicitly validating the effectiveness of the CNN-based calibrator. When the AoA of the UE is switched to $-15^{\circ}$, $15^{\circ}$ and $45^{\circ}$, the consistent improvement in the spatial spectrum and AoA estimation results across all cases is corroborated by the additional findings presented in Fig.~\ref{fig:figureSM2}e. Collectively, these results provide visual evidence that the CNN in the MoD-DNN method can effectively rectify the spatial spectrum output from coarray DBF, which in turn enhances AoA estimation performance through the sparse spatial spectrum obtained by the SCG algorithm.


\subsection{Performance of AoA estimator}

To enable a rigorous quantitative assessment and comparative analysis of estimation performance among various methods, we present a comprehensive statistical evaluation of the proposed MoD-DNN algorithm. In particular, we benchmark its performance against the MUSIC algorithm and the DeepMUSIC algorithm \cite{Elbir2020DeepMUSIC}. Furthermore, we extend our inquiry to include a purely data-driven CNN \cite{Wang20212DCNN}. Our prior investigations have explored different data-driven backbones \cite{Wang20212DCNN, Liu20235GNR} for similar tasks. However, in the context of this research, CNN is selected as a representative method for a more intuitive comparative analysis. Drawing parallels to the proposed MoD-DNN, it is noteworthy that the DeepMUSIC algorithm also produces a spatial spectrum, whereas CNN directly yields AoA estimates.

Our investigation begins by examining the root-mean-square-error (RMSE) a function of the SNR. As shown in Fig.~\ref{fig:figureSM2}f, both the MUSIC and DeepMUSIC methods exhibit relatively consistent performance levels despite variations in SNR. The saturated floors of the RMSE performance exhibited by MUSIC and DeepMUSIC can be attributed to their reliance on the ideal array response model. Due to the lack of specific calibration for angular-dependent errors, both MUSIC and DeepMUSIC experience performance degradation resulting from basis mismatch. In contrast, both the CNN and the proposed MoD-DNN demonstrate decreasing RMSEs as the SNR increases, with MoD-DNN consistently outperforming the CNN. This notable superiority stems from the effective utilization of signal model knowledge by MoD-DNN, which sets it apart from the purely data-driven CNN approach. Consequently, we observe that the advantage of MoD-DNN becomes more pronounced in the high SNR region.

To further validate the effectiveness of the proposed method in impairment calibration, we conduct a comparative analysis of RMSEs against the degree of impairment. In order to simulate varying degrees of impairment, we employ a weight coefficient $\rho$ ranging from $0$ to $1$ to weight the error matrix $\boldsymbol{\Gamma}$.  As such, the received signal in CFR form is adjusted as
\begin{align}\label{angular_dependent_error_1}
	\mathbf{h}(k) &= \sum\nolimits_{l=1}^{L}\rho\boldsymbol{\gamma}_k(\theta_l)\odot\mathbf{a}(\theta_l,d)\bar{s}_l(k)+\bar{\mathbf{n}}(k)=\rho\boldsymbol{\Gamma}_k\odot\mathbf{A}(\boldsymbol{\theta},d)\bar{\mathbf{s}}(k)+\bar{\mathbf{n}}(k).
\end{align}
When $\rho = 0$, no hardware impairment is contained in the received signal. As for $\rho=1$, the received signal is severely influenced by the hardware impairment depicted in Fig.~\ref{fig:figureSM1}.  As plotted in Fig.~\ref{fig:figureSM2}g, both DeepMUSIC and MUSIC exhibit favorable performance under low impairment conditions. However, their performances degrade as $\rho$ increases. This phenomenon suggests that the algorithms are influenced by hardware impairments. Conversely, the CNN and MoD-DNN algorithms demonstrate relatively stable performance with minimal fluctuations, which indicates that the methods are robust to hardware impairments. Additionally, by virtue of the integration of the CNN and SCG algorithm, MoD-DNN effectively mitigates the influence of hardware impairments, surpassing the purely data-driven CNN.

Then, we assess the convergence speed of different algorithms by examining the standard deviation (SD) of the MSE loss with respect to epoch. It is noteworthy that we take the convergence value as the expectation for calculating the standard deviation. As demonstrated in Fig.~\ref{fig:figureSM2}h, the SD of MoD-DNN converges rapidly compared to the CNN and DeepMUSIC, thereby showcasing the training efficiency advantage of MoD-DNN. The ability of fast convergence primarily stems from the incorporation of model knowledge to guide the learning process and the utilization of CNN parameter sharing strategy in the MoDL-SSR module, which significantly reduces the number of parameters to be learned.

In the final phase of our simulations, we examine the RMSE performance of the proposed MoD-DNN framework under varying channel conditions. Concretely, we consider $6$ different channels as sipulated in the 3GPP TR 38.901 standard, i.e.,  3GPP\_38.901 \_InH\_LoS, 3GPP\_38.901 \_InF\_LoS,  3GPP\_38.901 \_Umi\_LoS, 3GPP\_38.901\_Uma\_LoS and 3GPP\_38.901\_Rma\_LoS. As shown in Fig.~\ref{fig:figureSM2}i, MoD-DNN remains effective across these channel scenarios, with RMSE varying between $0.1^\circ$ and $1^\circ$. Compared to the outdoor scenarios, the superior performance of MoD-DNN in indoor scenarios, such as InF and InH, validate the suitability of the proposed MoD-DNN method for AoA estimation using 5G gNodeB.

\subsection{Computational Complexity}

As the final part of our simulation evaluation, we compare the computational efficiency of MUSIC, DeepMUSIC, CNN, and MoD-DNN. First, we analyze the time complexity of the MoD-DNN framework and other methods. As mentioned earlier, the MoD-DNN architecture consists of three main modules: the multi-task autoencoder module, the spatial spectrum generation module, and the MoDL-SSR module. For the first module, the multi-task autoencoder, the computational complexity is mainly influenced by the number of antennas ($N$), the number of subcarriers ($K$), and the number of hidden units in the feedforward network ($H$), leading to a complexity of $\mathcal{O}_1(N^2KH)$. In the second module, the spatial spectrum generation, the complexity is determined by the number of antennas and the size of the spatial spectrum ($M$), resulting in $\mathcal{O}_2(N^2M)$. The MoDL-SSR module, which comprises both model-driven and data-driven components, has its complexity split between the size of the spatial spectrum ($M$) for the model-driven part (SCG algorithm) and the channel size and kernel size ($C$) for the data-driven component (CNN). Thus, the complexity for this module is $\mathcal{O}_3(M^2 + C^2ML)$. In total, the overall complexity of the MoD-DNN framework can be expressed as $\mathcal{O}(N^2KH + N^2M + C^2ML)$. Similarly, we analyze the time complexity and computational load of other comparable algorithms, as shown in Table~\ref{table3}. The computational load is measured using floating point operations, and its unit is floating point operations per second (FLOPS).

\begin{table}[htbp!]
	\centering
	\caption{Time Complexity and Computational Amount for Different Algorithms}
	\label{table3}
	\begin{tabular}{c c c c c}
		\toprule
				& & & & \\[-9pt]
		&MUSIC&DeepMUSIC&CNN&MoD-DNN\\
		\hline
				& & & & \\[-6pt]
		\multirow{2}{*}{time complexity} &{$\mathcal{O}(KN^2$}& {$\mathcal{O}(N^2M$} & \multirow{2}{*}{$\mathcal{O}(C^2N^2KL)$} & $\mathcal{O}(N^2KH+$\\
		&$N^3+MN^2)$&$+C^2M^2L)$&&$+N^2M+C^2ML)$ \\
				& & & & \\[-6pt]
		computational load&\multirow{2}{*}{0.0002} &\multirow{2}{*}{0.213} &\multirow{2}{*}{0.017} &\multirow{2}{*}{0.134}\\
		$[$GFLOPS$]$&&&& \\
		\bottomrule
	\end{tabular}
\end{table}

While MUSIC exhibits lower computational requirements, the overall computational complexity of the other algorithms is of a similar order of magnitude. Runtimes were recorded using PyTorch's built-in timer on a PC with an Intel(R) Core(TM) i7-1065G7 CPU and 16 GB of RAM. The results are presented in Table~\ref{table4}, with training times specified in hours and testing times provided in milliseconds. Among the algorithms generating spatial spectra, CNN shows a significantly shorter training time, owing to its direct output of low-dimensional AoA values. Due to Python’s ability to leverage low-level libraries that optimize hardware performance for running neural networks, MUSIC does not demonstrate a notable advantage in runtime. In contrast, by using the weight-sharing strategy within the CNN module and the closed-form solutions provided by the SCG algorithm, MoD-DNN enjoys a reduction in the number of training parameters compared to DeepMUSIC, which consequently translates to a discernible advantage in training time. As for the testing time, all algorithms showcase millisecond-level computation speeds, which ensures real-time feasibility of implementation.

\begin{table}[htbp!]
\centering
\caption{Training and Testing Times for Different Algorithms}
\label{table4}
\begin{tabular}{c c c c c}
\toprule
	& & & & \\[-9pt]
	&MUSIC&DeepMUSIC&CNN&MoD-DNN\\
	\hline
	& & & & \\[-6pt]
	training [h]&-& 30.8 & 1.6 & 15.6 \\
	& & & & \\[-6pt]
	testing [ms]&9.4 &31.6 &1.8 &14.9\\
\bottomrule
\end{tabular}
\end{table}

Thus far, the numerical simulations have successfully validated the effectiveness of the proposed method in terms of impairment calibration, as well as its ability to improve estimation accuracy, training and testing efficiency. In terms of practical implementation, a standard commercial 5G BBU compatible with the 3GPP specification can process BB UL-SRS signals and output CFR measurements, as we will demonstrate in the subsequent real-world test. These measurements are then forwarded to the positioning server for final computation and resolution. As highlighted in this subsection, the computational load is extremely low and the real-timeness can be guaranteed with off-the-shelf 5G equipment.

\section{Experiment Validation and Analyses}
\label{sec:experiment}

\subsection{Experiment in an Anechoic Chamber}

\begin{figure*}[!ht]
\centerline{\includegraphics[width=1\linewidth]{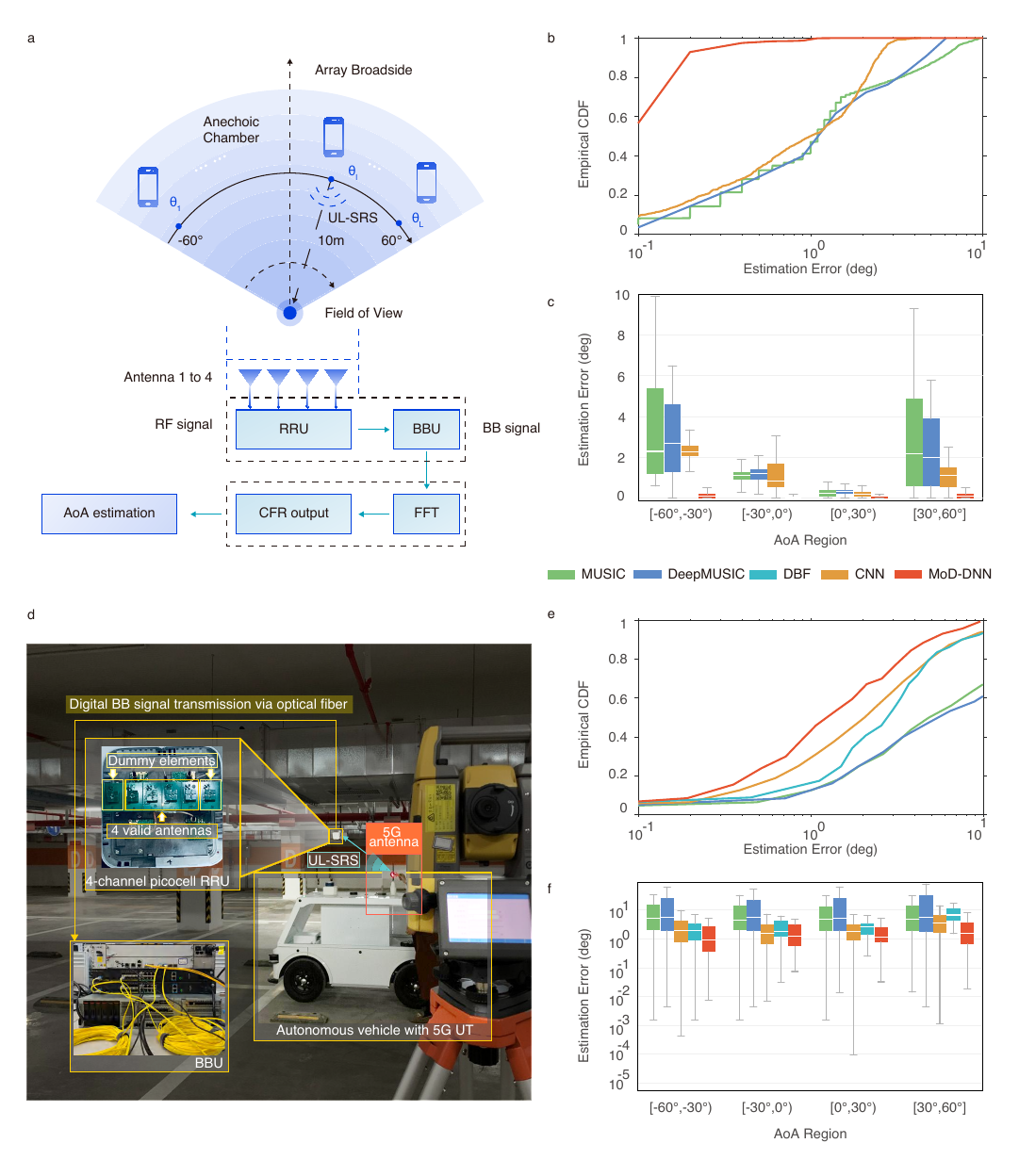}}
\caption{Experiment in an anechoic chamber and real-world test in an indoor environment.  \textbf{a}, Setups for anechoic chamber experiments. Error performance of different methods in an anechoic chamber in terms of  \textbf{b}, CDF and  \textbf{c}, boxplots for 4 subregions. \textbf{d}, Setups for  real-world test in an indoor experiment. Real-world error performance of different methods in terms of  \textbf{e}, CDF and  \textbf{f}, boxplots for 4 subregions.}
\label{fig:Figure3}
\end{figure*}

In this section, we present a comparative analysis of various methods for AoA estimation within an anechoic chamber setting. The experimental setups are visually depicted in Fig.~\ref{fig:Figure3}a. To mimic the 5G infrastructure within the chamber, we employ a 5G UE and a gNB for signal transmission and reception, respectively. During the data acquisition process, the AoA of the 5G UE is incrementally rotated from $-60^{\circ}$ to $-60^{\circ}$, with a uniform interval of $1^{\circ}$. For each AoA, a total of $450$ SRS symbols are transmitted by the 5G UE. The sounded CFRs are then captured by the gNodeB, yielding $121\times450 = 54450$ groups of CSI data. We extract $121\times400 = 48400$ groups of data for network training and reserve $121\times50 = 6050$ groups for validation purposes from the collected CSI data.\footnote{The parameter settings are consistent with those employed throughout the simulations.} We commence our analysis by examining the cumulative distribution function (CDF) curves of the estimation errors, as plotted in Fig.~\ref{fig:Figure3}b. Noteworthy to underscore is the controlled setting of the anechoic chamber, where the presence of non-line-of-sight (NLoS) conditions and the intricate dynamics of multipath propagation are effectively precluded. Consequently, the foremost challenge encountered in the accurate estimation of AoA predominantly emanates from hardware impairments. Our findings corroborate the superiority of the proposed MoD-DNN approach over alternative approaches with regard to accuracy benchmarks. The remarkable efficacy of the MoD-DNN technique becomes evident in its demonstrated capability to achieve a profound reduction of over $95\%$ in the $80^{\text{th}}$ percentile of estimation error, decreasing it from approximately $3^{\circ}$ to $0.15^{\circ}$. This in turn amplifies the method's efficacy in the critical domain of hardware impairment calibration.

In order to observe the AoA estimation performance of the MoD-DNN method more specifically, we proceed with the assessment of the AoA estimation performance of different methods by analyzing the corresponding boxplots of estimation errors across different angular regions. Specifically, the FoV is divided into four subregions: $[-60^{\circ},-30^{\circ})$, $[-30^{\circ},-0^{\circ})$, $[0^{\circ},30^{\circ})$ and $[30^{\circ},60^{\circ}]$. The upper and lower whiskers of the boxplots respectively signify the maximum and minimum estimation errors, while the bottom and top edges of the box indicate the first quartile (Q1) and third quartile (Q3), respectively. The horizontal line within the box represents the median. It is important to note that data points with errors $1.5$ times greater than Q3 are considered outliers and are not included in the presented results. As displayed in Fig.~\ref{fig:Figure3}c, within the narrower angular regions of $[-30^{\circ},-0^{\circ})$ and $[0^{\circ},30^{\circ})$, the estimation results of all algorithms exhibit relatively high accuracy, with median errors of approximately less than $1^{\circ}$. This is primarily attributed to the insignificant impact of angular-dependent phase errors within these regions. On the other hand, the curves of phase errors exhibit mild fluctuations, resulting in a low interquartile range (IQR) for the boxplots. Notably, the proposed MoD-DNN algorithm significantly outperforms the other methods across all evaluated metrics in the boxplots, including maximum error, median error, and IQR. This phenomenon underscores the robust estimation performance of MoD-DNN. As the AoA of UE extends to the wider angular spans of $[-60^{\circ},-30^{\circ})$ and $[30^{\circ},60^{\circ}]$, the phase errors are conspicuously exacerbated, which leads to a notable deterioration in the performance of both the MUSIC and DeepMUSIC algorithms. Specifically, their maximum errors soar to approximately $10^{\circ}$ and $6.5^{\circ}$, respectively, while their medians hover around $2^{\circ}$. The substantial fluctuation in phase errors also contributes to high IQR values for MUSIC and DeepMUSIC. In comparison, the CNN and MoD-DNN methods are less affected, while the maximum IQR of the CNN and MoD-DNN methods are respectively $1.42^{\circ}$ and $0.2^{\circ}$. This observation suggests that MoD-DNN offers a substantial improvement of up to $85.9\%$ in estimation stability compared to the purely data-driven CNN method.

The results substantiate the efficacy of integrating data-driven methodologies to glean insights into the hardware impairments inherent within the gNodeB. Through the strategic fusion of a priori model knowledge, MoD-DNN garners enhanced IQR and maximum error metrics compared to its pure data-driven CNN counterpart. This empirical validation underscores MoD-DNN's remarkable prowess in automatic calibration, thereby further establishing its effectiveness in real-world scenarios. Through the analysis of boxplots, we also gain insights into the performance of different methods in terms of AoA estimation across distinct angular regions. The superior performance of MoD-DNN, especially in the presence of severe phase errors associated with larger AoA values, is evident, thereby reinforcing its robustness and effectiveness in AoA estimation tasks.

\subsection{Real-world Field Test in an Indoor Environment}

To provide additional validation of the proposed MoD-DNN method and evaluate its localization performance in a complex and realistic environment, we conducted experiments in an underground parking lot. Fig.~\ref{fig:Figure3}d provides a photograph of the experimental environment and depicts the setup of the measurement system. The underground parking lot is characterized by the presence of numerous metallic plumbing pipes, poles, and thick pillars, which introduce intricate multipath effects to the wireless signals. By mounting a 5G user terminal (UT) atop an autonomous vehicle, we established an experimental configuration that allowed for automatic data collection. In conjunction with the 5G device, the vehicle was outfitted with a diverse array of sensors to precisely measure and record the UT locations. This comprehensive sensor suite enabled the acquisition of reliable ground truth data, which serves as a vital basis for conducting meticulous comparisons and in-depth performance evaluations. For technical details pertaining to the devices employed in the experiments, we refer interested readers to reference \cite{Pan2023In}.

In the experiment, we placed the remote radio unit (RRU) at a fixed position $(-10.3\; \mathrm{m}, 9.5\; \mathrm{m})$. The autonomous vehicle traveled within a rectangular area of approximately $1125\; {\mathrm{m}}^2$, bounded by the coordinates $[-35\; \mathrm{m},10\; \mathrm{m}]$ and $[8\; \mathrm{m},33\; \mathrm{m}]$. During its journey, the vehicle made stops at $476$ coordinates within the parking lot, with intervals ranging from $1.5$--$2.5\;\mathrm{m}$between each stop. We deliberately selected positions where the AoA fell within the range of $[-60^{\circ}, 60^{\circ}]$. As such, we collected CSI at a total of $279$ positions are used for performance evaluation. At each position, the 5G UT sent $100$ UL-SRS symbols, amounting to a dataset comprising $279\times100 = 27900$ groups of samples. To ensure a well-distributed validation set, we sorted the samples based on their true AoA values in ascending order. From every four consecutive positions, we extracted one position to form the validation set, while the remaining positions were used for network training. Given that the MUSIC algorithm is incompetent to estimate coherent multipath signals, we further included DBF algorithm as a comparative benchmark against the proposed MoD-DNN method, allowing us to highlight the strengths of MoD-DNN in terms of its calibration ability. The AoA estimation performance of all methods, including MoD-DNN, DBF, and others, are provided in Figs.~\ref{fig:Figure3}e and f. Fig.~\ref{fig:Figure3}e shows the empirical CDF curves of five AoA estimation methods. We readily find a discernible decline in performance for both the MUSIC and DeepMUSIC algorithms, in stark contrast to the outcomes demonstrated in Fig.~\ref{fig:Figure3}b. This performance decline can be ascribed to the rank-deficiency problems arising from NLoS and multipath propagation phenomena that are prevalent within indoor environment. Neither method is effective in resolving the resultant coherent signals, leading to errors exceeding $10^{\circ}$ at the $80{\rm{th}}$ percentile. In contrast, the DBF algorithm, in conjunction with the preprocessed coarray spatial spectrum, displays resilience against signal coherence. The CDF curve for DBF underscores its capability to offer reasonably accurate AoA estimates for the 5G UT, albeit slightly trailing behind the CNN method. Leveraging the LoS information present in the coarray DBF spectrum, the CNN within the MoD-DNN framework remains capable of mitigating phase errors, thereby enabling the SCG algorithm to produce the corresponding sparse spectrum. Consequently, the proposed MoD-DNN consistently outperforms alternative methods in indoor scenarios, showcasing its competence in overcoming both hardware impairments and multipath propagation, thereby improving AoA estimation accuracy.

In scenarios where the UT is situated in an NLoS position, the coarray spatial spectrum inherently conveys limited information concerning the true AoA. Consequently, the model-driven network's performance becomes constrained when relying on such spectra. Nonetheless, even in these challenging circumstances, the MoD-DNN achieves a commendable $80{\rm{th}}$ percentile error of $3.19^{\circ}$. This stands in stark contrast to the approximately $4.9^\circ$ error exhibited by CNN and DBF, marking a significant reduction of more than $34.9\%$. This observation clearly substantiates the advantageous attributes of the MoD-DNN, particularly in scenarios characterized by NLoS propagation. For a more detailed error analysis, we turn our attention to boxplots that represent different subregions, as depicted in Fig.~\ref{fig:Figure3}f. It is worth noting that due to the large error fluctuations observed in purely-model-driven methods, we opt for a logarithmic coordinate system to enhance the clarity of the boxplots. Despite the boxplots appearing relatively close in height on the graph, the IQR of the MoD-DNN is significantly lower than that of the comparison methods. To provide a specific instance, in the subregion $[-60^\circ,-30^\circ)$, the maximum IQR for MoD-DNN is $2.37^\circ$, whereas the minimum IQR within the same subregion for other methods is $3.1^\circ$ for DBF. This signifies a substantial $23.2\%$ enhancement in estimation accuracy stability. Additionally, under the combined influence of NLoS, multipath propagation, and hardware impairment, the differences between subregions are relatively less pronounced compared to the boxplots in Fig.~\ref{fig:Figure3}c. Nevertheless, the MoD-DNN algorithm consistently outperforms the other methods, manifesting as error reductions of at least $11.8\%$, $17.8\%$, and $20.1\%$ in Q1, Q3, and the median, respectively. These outcomes confirm that the systematic non-ideal factors, such as hardware impairments, multipath propagation, and NLoS conditions, exert detrimental effects on the AoA estimation process and negate the applicability of the conventional methods. Importantly, the MoD-DNN effectively mitigates these challenges, thereby enhancing overall estimation performance.

\section{Conclusion}
\label{sec:conclusion}

In this research endeavor, we present the MoD-DNN framework for accurate AoA estimation in the presence of hardware impairments in 5G gNodeBs. We leverage the complementary strengths of model-driven and data-driven methodologies and enhance the angular estimation performance in real-world scenarios characterized by the presence of hardware impairments. The proposed framework comprises multiple interconnected components that target specific challenges in the AoA estimation process. To enhance the consistency of the received signals, we first design a set of beamformers based on frequency diverse multi-task autoencoder, which effectively filters the signals into distinct subregions. On this basis, we overcome the adverse effects of angular-dependent phase errors by reformulating the AoA estimation task as an inverse problem of the spatial spectrum. A spectrum calibrator based on CNN is devised and we also introduce the SCG algorithm to derive closed-form solutions for the reconstruction of the sparse spatial spectrum. The proposed framework incorporates an iterative optimization process that leverages the interplay between the CNN and SCG algorithms. This iterative approach facilitates the efficient calibration of the spatial spectrum, resulting in a noteworthy enhancement of AoA estimation accuracy. We substantiate the efficacy of the proposed framework through comprehensive numerical simulations and real-data experiments conducted in both an anechoic chamber and an underground parking lot. The results clearly confirm the improved performance of AoA estimation achieved by the proposed MoD-DNN framework. By introducing the MoD-DNN framework and its constituent components, this research brings to light how mobile networks and smart positioning can join forces for a more connected world.

As a key capabilities of the 5G-Advanced stage and one of the core visions of 6G, mobile positioning can add a notable bonus onto base stations and terminals, allowing for numerous new tasks previously inconceivable. Mobile positioning is not developed to replace GPS or other positioning modalities, but simply because the base stations, as ubiquitous infrastructure equipped with resources such as power supply, antenna feeder, and transmission facilities, are already there. If the positioning capability can be gained through a mere software upgrade, why not catch the opportunity? The methodologies and findings presented herein offer valuable insights and lay the foundation for future advancements in this domain.
	
\begin{acks}
A preliminary version of this article was presented at the 38-th AAAI Conference on Artificial Intelligence (AAAI'2024), Vancouver, BC, Canada \cite{Liu2024}.
\end{acks}

\bibliographystyle{ACM-Reference-Format}
\bibliography{reference}

\end{document}